\documentclass[journal,9.5pt,final]{IEEEtran}
\newlength{\flexwidth}
\usepackage[T1]{fontenc}
\usepackage{cite}
\usepackage{url}
\usepackage{amsmath}
\usepackage{amssymb}
\usepackage{amsthm}
\usepackage{empheq}
\usepackage[cmintegrals]{newtxmath}
\usepackage{tabularx}
\usepackage{multirow}
\usepackage{booktabs}
\usepackage{diagbox}
\usepackage[font=small]{caption}

\usepackage[labelformat=simple]{subcaption}

\usepackage{dblfloatfix}
\usepackage{graphicx}
\usepackage{epstopdf}
\usepackage{xcolor}
\usepackage{mathtools,amssymb,lipsum}
\usepackage[detect-weight=true, binary-units=true]{siunitx}
\usepackage[inline]{enumitem}
\usepackage{cuted}
\usepackage{blindtext}
\usepackage[boxruled,norelsize]{algorithm2e}
\usepackage{setspace}
\usepackage{float}
\usepackage{siunitx}

\newtheorem{remark}{\change{Remark}}

\newcommand{\expectation}[1]{\text{E}\left\{#1\right\}}

\newcommand{\transpose}[0]{^\text{T}}

\newcommand{\change}[1]{#1}

\makeatletter
\newcommand{\removelatexerror} {\let\@latex@error\@gobble}
\makeatother

\begin{document}

\title{Multiservice-based Network Slicing Orchestration with Impatient Tenants}

\author{Bin~Han, ~\IEEEmembership{Member,~IEEE,}
	    Vincenzo~Sciancalepore, ~\IEEEmembership{Senior Member,~IEEE,}\\
   	    Xavier~Costa-P\'erez, ~\IEEEmembership{Senior Member,~IEEE,}
   	    Di~Feng, and~Hans~D.~Schotten,~\IEEEmembership{Member,~IEEE}
\thanks{\textit{Bin Han and Hans D. Schotten are with Institute of Wireless Communication, Technische Universit\"at Kaiserslautern, 67655 Kaiserslautern, Germany. Emails: \{binhan, schotten\}@eit.uni-kl.de}}%
	\thanks{\textit{V. Sciancalepore and X. Costa-P\'erez are with NEC Laboratories Europe, 69115 Heidelberg, Germany. Emails: \{vincenzo.sciancalepore, xavier.costa\}@neclab.eu}}%
	\thanks{\textit{D. Feng is with Faculty of Business and Economics, Universit\'e de Lausanne, CH-1015 Lausanne, Switzerland. Email: di.feng@unil.ch}}%
}

\maketitle

\thispagestyle{empty}	
\begin{abstract}
The combination of recent emerging technologies such as network function virtualization (NFV) and network programmability (SDN) gave birth to the novel Network Slicing paradigm. 5G networks consist of multi-tenant infrastructures capable of offering leased network ``slices'' to new customers (e.g., vertical industries) enabling a new telecom business model: Slice-as-a-Service (SlaaS). However, as the service demand gets increasingly dense, slice requests congestion may occur leading to undesired waiting periods. \change{This may turn into impatient tenant behaviors that increase potential loss of the business attractiveness to customers}. In this paper, we aim to $i$) study the slicing admission control problem by means of a multi-queuing system for heterogeneous tenant requests, $ii$) derive its statistical behavior model, $iii$) find out the rational strategy of impatient tenants waiting in queue-based slice admission control systems, $iv$) prove mathematically and empirically the benefits of allowing infrastructure providers to share its information with the upcoming tenants, and $v$) provide a utility model for network slices admission optimization. Our results analyze the capability of the proposed SlaaS system to be approximately Markovian and evaluate its performance as compared to a baseline solution.

\end{abstract}


%
\IEEEpeerreviewmaketitle

\section{Introduction}


\emph{Network Slicing}~\cite{intro_slic}
is an emerging 5G technology that allows infrastructure providers to offer ``slices'' of resources (computational, storage and networking) to network tenants. In this way a new business game~\cite{networkslic}
is introduced as infrastructure providers (sellers) strategically decide which tenants (buyers) get granted slices to deliver their services. Intuitively, this involves a number of challenges that fall in the economic research field, which, in turn, requires a detailed understanding of the context. In particular, the infrastructure provider may rely on this emerging technology as a means to increase its revenue sources. However, to achieve the overall revenue maximization, advanced admission control policies are required as tenants compete for a limited set of available resources.


In this competing environment, a brokering solution may act as a mediator between seller and buyers while providing service level agreements (SLAs) guarantees to granted running slices~\cite{samdanis2016network}. 
Admission control policies will guide the broker in the process of deciding the set of network slices that can be installed on the system and the ones to be rejected. As the number of network slices grows---as envisioned for the next few years~\cite{slicMobicom2018}---it will be necessary to design an automated solution that dynamically decides on the received slice requests while guaranteeing a certain degree of fairness among network tenants. 
Indeed, network slice requests may be queued while waiting for the next available resources, or may be re-issued. 

To properly design such a \emph{slicing brokering process}, a deep understanding of the slice queuing behavior is needed that accounts, e.g. the average slice duration (based on the slice type), the frequency of slice requests (based on the tenant), etc. This enables a Slice-as-a-Service (SlaaS)~\cite{sciancalepore2017slice} solution that fully supports on-demand slices requests: tenants issue slice requests for given periods of time and decide whether to re-issue the same request upon rejection based on service level agreements. Advanced slicing admission control solutions may have different policies for tenants frequently asking for short-term slices---such as Internet-of-Things (IoT), or crowded event-based network slices---as they will automatically re-issue the same request in the near future, with respect to those that require only few longer network slices---such as Mobile Virtual Network Operators (MVNOs) or Industrial Network Slices~\cite{TII2018_slicing}---which may be probably lost if not accepted. 
Moreover, similar as widely recognized in all kinds of queuing systems for service scheduling, tenants may be \emph{impatient} and choose to leave for another available infrastructure provider instead of waiting in a queue, especially when the expected waiting time is long. \change{This relies on the reasonable assumption that multiple network operators may offer similar slice services over the same spatial area opening up to new business models as per the SlaaS concept. A concrete example could be a stadium area covered by multiple telco operators, where a short-term crowded event may require a tenant to ask for a slice that, in turn, can be offered by different operators based on different offers. 
Such impatient behavior shall also be taken into account while designing a slicing admission control solution to mitigate potential revenues loss in case of resource congestion.}

While conventional admission control problems have been extensively studied in the literature, \change{as our main contribution in this study, we pioneer a new stochastic model for network slicing that leverages on the multi-queuing system to optimally design an admission control of on-demand network slices as well as to orchestrate them once accepted. This also allows to account for impatient tenant behaviors and heterogeneous network slice characteristics while, at the same time, enforcing given performance metrics, such as fairness between tenants or between network slice types or utility-based maximization. We also numerically demonstrate this multi-queuing framework to outperform conventional single-queue solution.}


\change{The remainder of this article is structured as follows. In Section~\ref{sect:model}, we introduce our assumptions and we formulate the network slicing admission control problem for on-demand slice request arrivals. In Section~\ref{sect:ns_queue}, we provide a simple use case to cast our problem into a queuing system. In Section~\ref{sect:multi-queue-controller}, we model the problem as a multi-queue problem where each queue may host slice requests of the same type while waiting for being granted. In Section~\ref{sect:controller}, we devise and analyze the multi-queuing controller with additional metrics by proving the capability of conventional queuing models on such system with impatient tenants taken into account, whereas in Section~\ref{sect:rational_impatient_tenants} we further deepen our analysis on the rational impatient behavior of tenants. In Section~\ref{sect:optimization} we briefly introduce a scheme to optimize the MNO's admission control strategy. In Section~\ref{sect:perf_eval}, we carry out an exhaustive simulation campaign to prove our findings and validate our model. In Section~\ref{sect:discussions} we discuss the applicability of some assumptions whereas in Section~\ref{sect:rel_work} we outline the main related works on this topic. Finally, in Section~\ref{sect:concl} we provide concluding remarks.}
\begin{table*}[!htbp]
	\centering
	\caption{\change{Key Notations}}
	\label{tab:notations}
		\begin{tabular}{l l}
			\toprule[2px]
			\textbf{Notation}&\textbf{Meaning}\\
			\midrule[1.5px]
			$M,N,\mathcal{N}$&Amount of resource types, amount of slice types, set of slice types\\\hline
			$\mathbf{r},\mathbf{s},\mathbf{a}$&Resource pool, set of active slices, assigned resources\\\hline
			$\mathbf{C},\mathbf{c}_n$&Resource cost matrix, resource cost of a type-$n$ slice\\\hline
			$\mathbb{S},\mathbb{A}$&Feasibility region, admissibility region\\\hline
			$\Delta\mathbf{s}_n$&unit slice incremental vector of type-$n$\\\hline
			$\mathbf{\Phi}$, $\Phi_i$&MNO's preference matrix, MNO's preference vector in state $\mathbf{s}_{(i)}$\\\hline
			$l_n, L_n, \overline{W}_n,$&Current queue length, average length, and average waiting time in queue $n$\\\hline
			$\lambda_n,\mu_n,\rho_n$&Request arrival rate, serving rate, and work load rate in queue $n$\\\hline
			$b_n,\beta_n$&Tenants' balking rate and balking exponent in queue $n$\\\hline
			$W_{\max{},n},\alpha_n$&Tenant's maximal waiting time and reneging exponent in queue $n$\\\hline
			$u_{0,n},u_n$&Tenants' one-time cost to issue a request and periodical cost to wait in queue $n$\\\hline
			$\zeta_n, \tau_n$&A tenant's periodical profit from a type-$n$ slice, the slice's random lifetime\\\hline
			$\eta_n$&Releasing rate of type-$n$ slices\\
			$\ni$&Maximal waiting cost w.r.t. expected achievable profit of a blind reneging tenant\\\hline
			$\Delta K$&Minimal waiting time of a tenant only aware of its position in queue\\
			\bottomrule[2px]
	\end{tabular}
\end{table*}

\section{Model design}
\label{sect:model}
We cast our problem into a typical network slicing scenario, where the Mobile Network Operator (MNO) decides to lease infrastructure resources to tenants, willing to pay to take over the control of an independent network slice so as to deliver an end-service to their own users.
Hereafter, we describe our assumptions and mathematically formulate the problem.

\subsection{Resource pool and slice types}
Let us consider a single MNO that possesses a static resource pool of $M$ different resources and offers $N=|\mathcal{N}|$ pre-defined types of slices. Depending on the slice type $n\in\mathcal{N}$, it costs a certain resource bundle to create and maintain a slice.  Let $\mathbf{r}=[r_1,r_2,\dots,r_M]^\text{T}$, $\mathbf{s}=[s_1,s_2,\dots,s_N]^\text{T}$ and $\mathbf{c}_n=[c_{1,n},c_{2,n},\dots,c_{M,n}]^\text{T}$ denote the resource pool, the set of active slices \change{(i.e. slices created upon accepted requests),} and the resource bundle required to maintain a slice of type $n\in\mathcal{N}$, respectively. The assigned resources can be then represented as
\begin{equation}
\mathbf{a}\overset{\Delta}{=}[a_1,a_2,\dots,a_M]^\text{T}=\mathbf{C}\times\mathbf{s},
\end{equation}
where $\mathbf{C}=[\mathbf{c}_1,\mathbf{c}_2,\dots,\mathbf{c}_N]$. 
At any time instance, the MNO cannot simultaneously maintain more slices than its resources may support. This constraint is expressed by the \emph{feasibility region}\cite{han2018slice}:
\begin{equation}
\mathbb{S}=\{\mathbf{s}\vert r_m-a_m\ge 0,\quad\forall 1\le m\le M\}. 
\end{equation}
Note that $\mathbb{S}$ is a finite discrete set, \change{thus the MNO can be characterized as a finite-state machine (FSM)} where each active slice set represents the system state $\mathbf{s}\in\mathbb{S}$.

\subsection{Slice admission in SlaaS}
We consider a certain number of tenants randomly generating network slice requests. Slices requested by a certain tenant are of the same type. For each tenant, the inter-arrival time between two requests is drawn from an exponential distribution. The request arrivals of different tenants are independent and identically distributed (i.i.d.).

\begin{figure}[!hbtp]
	\centering
	\includegraphics[width=.9\flexwidth]{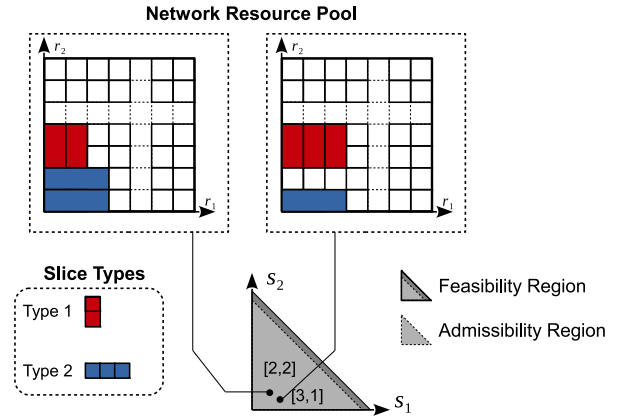}
	\caption{The network resource utilization can be described with the set of active slices $\mathbf{s}$, which always falls in the feasibility region $\mathbb{S}$. The admissibility region $\mathbb{A}$ is a proper subset of $\mathbb{S}$.}
	\label{fig:admissibility_region}
\end{figure}

Once a request for slice creation is triggered, the MNO makes a binary decision, i.e., the MNO either accepts or declines it. Upon acceptance, the requested slice is created, and continuously maintained so that a corresponding resource bundle is occupied until the slice is terminated (at the end of its lifetime) and the resource bundle is released. It should be noted that the constraint of feasibility region forbids the MNO to accept any request when its current state is close to the border of $\mathbb{S}$. In other words, if the current MNO resource pool is close to be saturated by active slices, it does not accept additional network slice requests that might experience a service disruption.
This introduces the well-known concept of \emph{admissibility region}\change{, where the idle resources are sufficient to allow to accept at least one new request\footnote{The admissibility region has been exhaustively studied in the literature for different use cases and scenarios. We refer the reader to~\cite{bega_TMC2019}, where a stochastic admissibility region is derived for a network slicing admission control.}:}
\begin{equation}
\mathbb{A}=\{\mathbf{s}\vert\mathbf{s}\in\mathbb{S},\exists n:\mathbf{s}+\Delta\mathbf{s}_n\in\mathbb{S}\},
\end{equation}
where $\Delta\mathbf{s}_n$ is the \emph{unit slice incremental vector} of type $n$
\begin{equation}\label{equ:slice_incremental_vector}
\Delta\mathbf{s}_n=[\underbrace{0,\dots,0}_{n-1},1,\underbrace{0,\dots,0}_{N-n}],\quad n\in\{1,2,\dots,N\}.
\end{equation}
Fig.~\ref{fig:admissibility_region} briefly illustrates the concepts of $\mathbb{S}$ and $\mathbb{A}$ with an example where $M=2, N=2$.

We assume that the lifetime of every slice is an i.i.d. exponentially distributed variable and the expected lifetime depends on the slice type. We also consider that the MNO makes every decision according to a consistent slicing policy, i.e., the decision depends only on the type of requested slice $n$ and the current system state $\mathbf{s}$ that defines the current set of active slices.

\subsection{Delayed reattempt upon request denial}
If a request for slice creation is declined---because of a temporary shortage of available resources due to many other active slices---the tenant is not able to obtain the requested slice immediately. Instead, its request may be sent to the MNO again for a reconsideration after some delay with the hope that some running slice has expired (i.e., resources have been released). Generally, there are two critical features of the delaying mechanism, which should be taken into account: $i$) resource efficiency and $ii$) fairness. The former requires that the chosen mechanism purses the resource pool utilization maximization whereas the latter requires that the expected delay for different requests is normalized.

Two categories of approaches are commonly used to solve this kind of problem:

\noindent\textbf{Random delay}. Every declined request is re-proposed to the MNO after a random delay. This approach provides a good fairness, but generates extra signaling overhead while being not able to provide the discipline of ``First Come, First Served'' (FCFS), as described in the next section.

\noindent\textbf{Queuing}. Declined requests wait in one or multiple queue(s) for the next opportunity during the MNO's decisional process. This is the most common solution in cloud service scheduling.

Hereafter, we show how a multi-queuing system may be fully exploited to provide insights on the system behaviors and pave the road towards a slicing orchestration solution.

%

\section{Network slicing queuing}
\label{sect:ns_queue}
In the literature a number of disciplines have been studied to serve the request queues. Among the others, the most common policies are $i$) First come, first served (FCFS), $ii$) Last come, first served (LCFS), $iii$) Random selection for service (RSS) and $iv$) Priority-based (PR). All of them analyze different behaviors and are used to achieve distinct performance metrics. For instance, the LCFS is used to \change{allow the latest arriving request to override its awaiting predecessors, such as in information-freshness-critical scenarios~\cite{yates2019age}. The PR is implemented when there is some high-level preference to be considered in centralized scheduling~\cite{khojasteh2019prioritization}.} RSS shows huge complexity in the implementation without bringing any significant advantage with respect to the others. Hereafter, \change{we discuss different queuing schemes with focus on the FCFS case}. However, any other discipline may be easily adapted to our analysis.
\subsection{Queuing schemes}
We differentiate the queuing systems into two different categories: $i$) single-queue and $ii$) multi-queue systems. When considering the single-queue, only one queue is implemented for all declined requests that need to wait for the next acceptance opportunity, an example was given in \cite{han2019markov}. Conversely, the multi-queue system implements multiple queue for declined requests. Specifically, such queues may show different features. We consider homogeneous-mixed queues, wherein each queue consists of requests for slices of different types, and heterogeneous queues, where each queue is specified for only one unique slice type. We next show a simple case-study to justify that the queuing system is suitable for this kind of problems.

\subsection{Resource efficiency: a simple case-study}
Consider a simplified case where $M=1$, $N=2$, $\mathbf{r}=[1]$, $\mathbf{c}_1=[0.6]$, $\mathbf{c}_2=[0.2]$ and $\mathbf{s}=[1,0]\transpose$. The first four requests awaiting in the queue(s) are in the sequential order $[1,1,2,2]$. The MNO takes a greedy strategy\change{, i.e., it intends to accept all requests so far the resource pool supports.}

\begin{figure}[!htbp]
	\centering
	\includegraphics[width=\flexwidth]{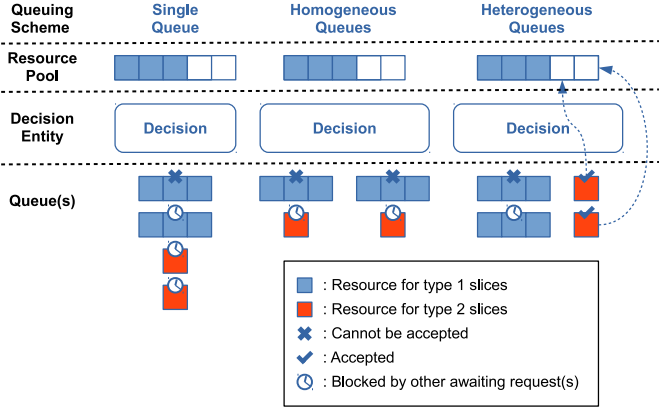}
	\caption{A simple case study on different queuing schemes.}
	\label{fig:schemes}
\end{figure}

\change{In both the schemes of single-queue and homogeneous multi-queue, the awaiting requests of type~$2$ are blocked from acceptance due to the type-$1$ requests ahead of them, and therefore have to wait in queue, although the MNO has both enough idle resource and the intention to accept them immediately.} The heterogeneous multi-queue scheme, in contrast, enables the MNO to fully utilize its resources as shown in Fig.~\ref{fig:schemes}.

Obviously, both the single-queue and the homogeneous multi-queue schemes can also overcome this issue by introducing a ``queue-jumping'' mechanism. However, this may require an extra design of (more complex) logic that automatically (and dynamically) selects the queue jumper(s). Therefore, in this study we consider the scheme with $N$ FCFS heterogeneous queues. \change{Note that despite the intuitiveness and case-driven nature of this motivation, in Section VIII we demonstrate the performance superiority of this multi-queuing scheme through numerical simulations. Furthermore, as a well-studied extension of single-queue, the homogeneous multi-queue scheme is known to benefit only from its linearly increased serving rate in comparison to single-queue. Unfortunately, this gain does not apply in slice admission control where the request serving rates of different queues are jointly limited by the shared resource pool, while the implementation complexity is significantly higher than single-queue. Hence, in this study the performance of homogeneous multi-queue scheme will not be discussed.}

\section{Heterogeneous multi-queue admission control}
\label{sect:multi-queue-controller}
Based on the heterogeneous multi-queue scheme, we propose in this section a novel code to present the MNO's preference for different slice types in variable states, a multi-queue admission controller for SlaaS, and analyze its queue model.

\subsection{Slice-type preference encoder}
Differing from existing studies that do not consider queuing and the single-queue scheme, in the multi-queue scheme, the MNO may receive multiple requests for slices of different types \change{within the same operations period (which is usually synchronized to the billing cycle, e.g. a month or longer)}. Therefore, instead of making a binary decision to accept or decline a request, it has to either accept one among the simultaneously \change{awaiting requests} while declining the rest, or decline all of them\footnote{\change{Note that any simultaneous acceptance of multiple requests, which is technically applicable, can be decomposed into a sequence of atomic single-request acceptance operations. By considering only atomic acceptance operations, the action space is minimized.}}. Especially, with heterogeneous queues, the MNO's preference for some request queue(s) over the others implies its proclivity to some slice type(s) against the others.

\change{It shall be noted that the decision of accepting a request from a certain queue  requires the queue to be non-empty, which makes the MNO's decision dependent not only on the active slice set, but also on the status of queues. This extends the state space and increases the problem complexity. To simplify the analysis, we propose to investigate not the decision but the MNO's preference, which is decoupled from the queue status. However, it would be possible to uniquely determine the decision when any queue status is provided. Specifically, for an MNO offering} $N$ different slice types, we can encode an arbitrary preference of the MNO into a \emph{preference vector} of length $N+1$:
\begin{equation}
\change{\Phi=[\varphi_1,\varphi_2,\dots,\varphi_{N+1}]\transpose},
\end{equation}
which is a permutation of  $\{0,1,\dots,N\}$. \change{Every element $\varphi_i>0$ indicates a slice type and its position in $\Phi$ represents the MNO's preference. More specifically, $\forall i,j\in\{1,2,\dots N+1\}$, $i<j$ denotes that MNO prefers slice type $\varphi_i$ over $\varphi_j$. The zero-element denotes reserving resource for potential opportunities in future, so that all the requests of type $\varphi_j$ will not be served by the MNO at all, if $j>i$ and $\varphi_i=0$.}

While being in states on (or close to) the border of feasibility region $\mathbf{s}\in\mathbb{S}-\mathbb{A}$, the MNO cannot accept further request from any queue, hence the preference does not make any impact. Thus, we focus on the admissibility region $\mathbb{A}$ and assume that the MNO's preference is consistent and depends only on its current state $\mathbf{s}\in\mathbb{A}$. Thus, we can characterize the MNO's admission strategy with a $(N+1)\times\vert \mathbb{A}\vert$ \emph{preference matrix} as the following
\begin{equation}\label{equ:phi_matrix}
\begin{split}
\mathbf{\Phi}&=[\Phi_1,\Phi_2,\dots,\Phi_{\vert\mathbb{A}\vert}]
\\&
=\begin{bmatrix}
\phi_{1,1}&\phi_{1,2}&\dots&\phi_{1,\vert\mathbb{A}\vert}\\
\phi_{2,1}&\phi_{2,2}&\dots&\phi_{2,\vert\mathbb{A}\vert}\\
\vdots&\vdots&\ddots&\vdots\\
\phi_{N+1,1}&\phi_{N+1,2}&\dots&\phi_{N+1,\vert\mathbb{A}\vert}\\
\end{bmatrix},
\end{split}
\end{equation}
where each column $\Phi_i$ represents the preference for slice types in a specific \change{admissible state $\mathbf{s}_{(i)}\in\mathbb{A}$, for which the index $i$ is arbitrarily mapped by $\mathcal{I}:\mathbb{A}\overset{\mathcal{I}}{\rightarrow}\{1,2,\dots\vert\mathbb{A}\vert\}$.}

\subsection{Mechanism overview}\label{subsec:overall_mechanism}
\begin{figure}[!htbp]
	\centering
	\begin{subfigure}[b]{.95\flexwidth}
	\removelatexerror
	\begin{algorithm}[H]
	    \setstretch{0.9}
		\footnotesize
		Initialize with certain $N$, $\mathbb{S}$, $\mathbb{A}$, $\mathbf{\Phi}$ and $\mathbf{s}$\;
		\While(\hfill\emph{Main loop}){True}{
			Wait for the next incoming tenant issue\;
			\uIf(\hfill\emph{Releasing a slice}){Slice of type $n$ released}{$\mathbf{s}\gets\mathbf{s}-\Delta\mathbf{s}_n$\;}
			\ElseIf(\hfill\emph{Request arrives}){Slice of type $n$ requested}{
				$l_{n}\gets l_{n}+1$\;				
			}
			\While(\hfill\emph{Recursively serving the queues until blocked}){$\mathbf{s}\in\mathbb{A}$}{
				$\tilde{\mathbf{s}}\gets\mathbf{s}$\;
				Find the current preference vector $\Phi$ according to $\mathbf{\Phi}$ and $\mathbf{s}$\;
				\For(\hfill\emph{Serve queues w.r.t. preference}){$1\le n\le N$}{
					\uIf(\hfill\emph{Omitting queues after 0}){$\varphi_{n}=0$}{break\;}
					\ElseIf(\hfill\emph{Acceptance}){$l_{n}>0$ AND $\left(\mathbf{s}+\Delta\mathbf{s}_{n}\right)\in\mathbb{S}$}{
						$l_n\gets l_n-1$\;
						$\mathbf{s}\gets\mathbf{s}+\Delta\mathbf{s}_n$\;
					}
				}
				\If(\hfill\emph{Blockage detection}){$\tilde{\mathbf{s}}=\mathbf{s}$}{Break\;}
			}
		}	
	\end{algorithm}
	\caption{Pseudo code of the multi-queue slice admission controlling algorithm.}
	\label{fig:multi-queue_sac_algorithm}
	\end{subfigure}
	\begin{subfigure}[b]{.9\flexwidth}
		\centering
		\includegraphics[width=\columnwidth]{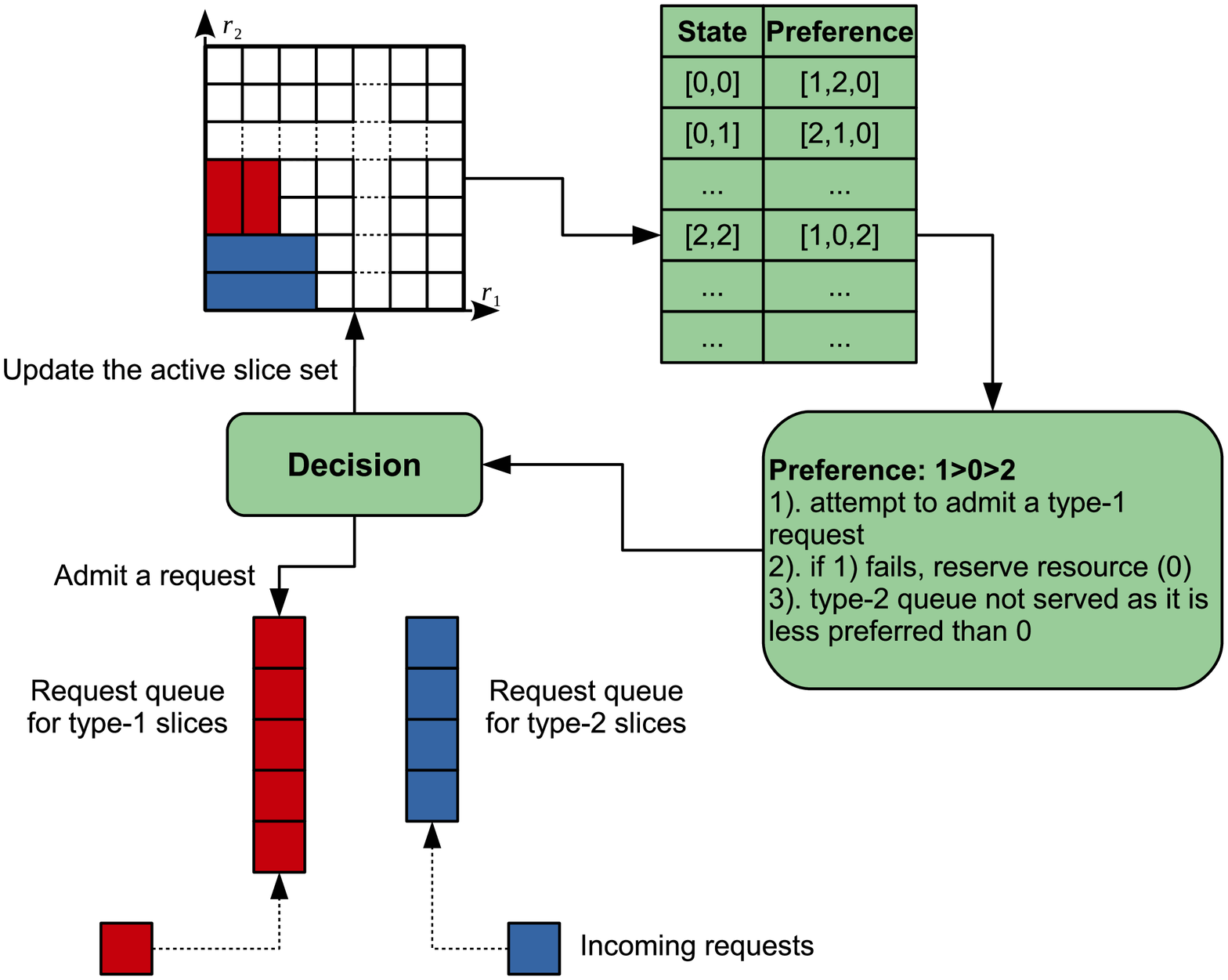}
		\caption{Graphical illustration.}
		\label{fig:multi-queue_sac_illustration}
	\end{subfigure}
	\caption{A heterogeneous multi-queue mechanism can be implemented to enable delayed slice admission, in which a specific preference among different slice types is addressed to every state of active slice set.}
	\label{fig:multi-queue_sac}
\end{figure}

Let $l_n$ denote the length of queue $n$, the decision entity executes the algorithm described in Fig. \ref{fig:multi-queue_sac}. The MNO keeps waiting for incoming tenant issues and responses upon issue arrivals. If the tenant issues to release a slice of its own, the MNO always releases it. \change{If the tenant issues to create a new slice, the request will be pushed into a queue with respect to the type of requested slice.} After responding to the issue, the MNO will recursively serve the request queues in a sequence determined by its admission strategy and active slice set, until no more waiting request can be accepted. Then it stops serving the queues and waits for the next tenant issue.

\section{Network slicing controller design}
\label{sect:controller}
Hereafter we analyze different characteristics of the conventional queuing models, highlighting the novel features applied to our model while designing the network slicing controller. This helps to shed light on the main advantages and limitations of our novel admission control model.

\subsection{Analysis of inter-acceptance time}
\change{As per existing works~\cite{bega2017optimising,han2018slice,bega_TMC2019}, we consider request arrivals of every slice type as an independent Poisson process, so that the inter-arrival time between requests in every queue is an independent exponential random process.} Conversely, the request acceptance rate of every queue is jointly determined by the slice releases of all types, and the MNO's preference strategy.

\begin{remark}\label{remark:geom_iat}
	Consider a heterogeneous multi-queue slice admission controller that executes the algorithm in Fig. \ref{fig:multi-queue_sac_algorithm} with a consistent preference matrix. The acceptance in different queues are mutually independent Poisson processes, if:	
	\begin{enumerate*}
		\item the arrivals of new requests and releases of active slices are mutually independent Poisson processes for every individual slice type;
		\item the arrivals of different slice types are mutually independent from each other, the releases of different slice types are mutually independent from each other.
	\end{enumerate*}\footnote{We refer the readers to \cite{han2019utility} for a detailed proof of \change{Remark}~\ref{remark:geom_iat}.}
\end{remark}

\subsection{Queuing-theoretic analysis}
While considering both request arrivals and request acceptances (service) as Poisson processes, every request queue is a classic $\text{M}/\text{M}/1$ queuing system, known as single-server birth-death system~\cite{shortle2018fundamentals}. Hence, many features of birth-death model can be directly applied.
\subsubsection{Little's Formula}
For slice type (queue) $n$, given its request arrival rate $\lambda_n$, according to the famous Little's formula\cite{little1961proof} the mean length of queue $n$ is
\begin{equation}\label{equ:little}
L_n=\lambda_n\overline{W}_n,
\end{equation}
where $\overline{W}_n$ represents the average waiting time in queue $n$.

\subsubsection{Steady Queue State Probability}
Given the request arrival \change{rate $\lambda_n$ and serving rate $\mu_n$ of queue $n$}, the probability that the queue steadily consists of $l$ requests at an arbitrary time instant is geometrically distributed, i.e.,
\begin{equation}\label{equ:steady_queue_state}
p_n(l)=(1-\rho)\rho^l,
\end{equation}
where $\rho_n=\lambda_n/\mu_n<1$ is the \emph{work load rate} of queue $n$.

\subsubsection{Waiting Time Distribution}
The probability density function (PDF) and the cumulative density function (CDF)  of an arbitrary type-$n$ request's waiting time are
\begin{align}
f(W_n)&=\begin{cases}
0& W_n<0\\
(\mu_n-\lambda_n)e^{-(\mu_n-\lambda_n)W_n}&W_n\ge 0
\end{cases},\label{equ:pdf_waiting_time}\\
F(W_n)&=\begin{cases}
0& W_n<0\\
1-e^{-(\mu_n-\lambda_n)W_n}&W_n\ge 0
\end{cases}.\label{equ:cdf_waiting_time}
\end{align}

\subsection{Extension: impatient tenants}
From Eqs. (\ref{equ:little}--\ref{equ:cdf_waiting_time}) it is clear that both $L_n$ and $W_n$ converge only when $\lambda_n<\mu_n$. Otherwise, when the request acceptance rate is below the arrival rate in queue $n$, the queue length will infinitely increase, and therefore also the mean waiting time. This is known as the necessary and sufficient condition of statistical equilibrium in queuing processes, as proven in \cite{kendall1951some}.

However, in a real slice admission controller, there are various situations where $\lambda_n\ge\mu_n$ for some $n$, including cases
\begin{itemize}
	\item when the controller is specified with an inappropriate strategy, so that requests in the queue $n$ is rarely or even never accepted despite of resource feasibility;
	\item when the release rates of active slices are low, so that the resource pool fails to support a sufficiently high $\mu_n$ regardless of any admission strategy.
\end{itemize}
There are two mechanisms that prevent queuing systems from such divergence.
On the one hand, the system may force to truncate a queue at some maximal length, and forbid this queue to take any new request before it is shortened. On the other hand, the clients may lose patience while waiting, and leave the queues before being served (e.g., for looking for some other MNO with resource availability). In the scenario of SlaaS, the system (MNO) is probably very cautious with refusing requests,  while the waiting time can be critical to the customers (tenants). Therefore, here we consider no queue truncation but queues with impatience.

Usually, impatience in queues can occur in three different behaviors:
$i$) balking, i.e. customers being reluctant to join a queue upon arrival, $ii$) reneging, i.e. customers leaving the queue after joining and waiting, and $iii$) jockeying from long lines to shorter ones. As the heterogeneous multi-queue design disables jockeying, here we consider the balking and reneging phenomena.

\noindent{\bf Balking Model.} The phenomenon of balking can be modeled in such a way, that every arrival request of slice type $n$ enters the queue with a probability $b_n$, which is a monotonically decreasing function of the current queue length $l_n$.
\emph{Ancker} and \emph{Gafarian} have proposed two different balking models in~\cite{ancker1963some1,ancker1963some2}. The first model considers a linear balking factor:
\begin{equation}
	1-b_n=l_n/l_{n,\max{}},
\end{equation}
where $l_{n,\max{}}$ is the truncation length of queue $n$. The second considers a non-linear balking factor:
\begin{equation}\label{equ:hyperbolic_balking_model}
1-b_n=\begin{cases}
0&l_n=0;\\
1-\beta_n/l_n&l_n\in\mathbb{N}^+,
\end{cases}
\end{equation}
where $\beta_n\in[0,1]$ measures the willingness of tenants requesting type-$n$ slices to wait. In cases that the tenant has knowledge about $\mu_n$, \emph{Shortle} et al. suggest another non-linear balking model~\cite{shortle2018fundamentals}:
\begin{equation}\label{equ:exp_balking_model}
	1-b_n=1-e^{-\beta_n l_n/\mu_n},\quad\beta_n>0.
\end{equation}

\noindent{\bf Reneging  Model.} The phenomenon of reneging can be modeled by randomly assigning an individual maximal waiting time to every request when it joins the queue. The request will leave the queue after that maximal waiting time if it has not been accepted yet. \emph{Ancker} and \emph{Gafarian} \cite{ancker1963some2} proposed to consider exponentially distributed random maximal waiting time:
\begin{equation}\label{equ:exp_reneging_model}
	W_{\max{},n}\sim\text{Exp}(\alpha_n),
\end{equation}
 where $1/\alpha_n>0$ is the mean maximal waiting time in queue $n$.

Here we consider the exponential balking and reneging models described by Eq.~\eqref{equ:exp_balking_model} and Eq.~\eqref{equ:exp_reneging_model}, respectively. We will justify this choice later in Section~\ref{sect:rational_impatient_tenants}. Before that, we first continue analyzing the performance of out heterogeneous multi-queue slice admission controller, taking into account the impatience of tenants.

\subsection{Performances with balking and reneging}\label{subsec:performances_with_balking_and_reneging}
It should be noted that the balking and reneging processes are with memory, leading to a non-Markovian behavior of request acceptances. However, under low balking and reneging rates, this impact can be negligible and the acceptance process can still be approximated as Poissonian. When the balking and reneging rates rise to significant levels, the memory of acceptance process shall be considered, as demonstrated in Section~\ref{subsec:geometric_iat} by means of simulations.

Under a combination of exponential balking and exponential reneging, the steady state probability of having $l$ requests in the queue $n$ is

\begin{equation}
p_n(l)=\begin{cases}
\left(1+\sum\nolimits_{j=1}^{+\infty}\frac{\Gamma(\gamma_n+1)(\frac{\lambda_n\delta_n^j}{\alpha_n})^l}{\Gamma(l+\gamma_n+1)}\right)^{-1}&l=0\\\\
p_n(0)\prod\nolimits_{j=1}^{l}\frac{\lambda_n\delta_n ^j}{\mu+j\alpha_n}&l\in\mathbb{N}^+
\end{cases},
\end{equation}
where $\gamma_n=\mu_n/\alpha_n$, $\delta_n=e^{-\beta_n/\mu_n}$ and  $\Gamma(\cdot)$ is the Gamma function. A detailed calculation is provided in the appendix.

Meanwhile, we are interested in three different distributions of waiting time spent in a queue $n$: $i$) $f_\text{a}(W_n)$ for requests that are eventually accepted, $ii$) $f_\text{r}(W_n)$ for requests that renege and $iii$) $f_\text{q}(W_n)$ for all requests that join the queue.
Let us define $A_n$ and $J_n$ as the events of request being accepted and joining the queue $n$, respectively. Following the analysis approach used in \cite{ancker1963some1,ancker1963some2}, there are 

\begin{align}
P(J_n)&=\sum\nolimits_{j=1}^{+\infty}p_n(j)\delta_n^j\\
P(A_n)&=p_n(0)+\sum\nolimits_{j=1}^{+\infty}{p_n(j)\delta_n^j\gamma_n}/{\left(\gamma_n+j\right)},\\
P(A_n,J_n)&=\sum\nolimits_{j=1}^{+\infty}{p_n(j)\delta_n^j\gamma_n}/{\left(\gamma_n+j\right)},\\
P(A_n\vert J_n)&={P(A_n,J_n)}/{P(J_n)}.
\end{align}
It can be obtained that
{
	\begin{align}
	f_\text{a}(W_n)=&\frac{p_n(0)\alpha_ne^{-(\mu_n+\alpha_n)W_n}}{P(A_n,J_n)}\sum\nolimits_{l=1}^{+\infty}\frac{\left(1-e^{-\alpha_n W_n}\right)^l\delta_n^\frac{l(l+1)}{2}}{l!(l-1)!}\\
	f_\text{r}(W_n)=&\alpha_ne^{-\alpha_nW_n}\frac{1-P(A_n\vert J_n)g(W_n)}{1-P(A_n\vert J_n)},\\
	f_\text{q}(W_n)=&P(A_n\vert J_n)\left[f_\text{a}(W_n)-\alpha_ne^{-\alpha_nW_n}g(W_n)\right]+\alpha_ne^{-\alpha_nW_n},
	\end{align}}
where $g(W_n)=\int_{0}^{W_n}e^{\alpha_n\xi}f_\text{a}(\xi)\text{d}\xi$.

The expectations of waiting times are therefore
	\begin{align}
	\overline{W}_{\text{a},n}&=\int_{0}^{+\infty}f_\text{a}(W_n)\text{d}W_n,\\
	\overline{W}_{\text{r},n}&={1}/{\alpha_n}-{P(A_n\vert J_n)\overline{W}_{\text{a},n}}/{\left[1-P(A_n\vert J_n)\right]},\\
	\overline{W}_{\text{q},n}&={\left[1-P(A_n\vert J_n)\right]}/{\alpha_n}.
	\end{align}

\section{Decision analysis of impatient tenants}\label{sect:rational_impatient_tenants}
To decide which models shall be used to describe the behavior of impatient tenants in the SlaaS scenario, we take the tenant's point of view and consider the business model of every individual tenant service instance through its life cycle. A rational strategy of impatience shall be obtained from the perspective of decision making by tenants.

\subsection{Tenant business model}\label{subsec:tenant_business_model}
Generally, the motivation of tenants to request network slices is to fulfill the end-user service demands from their own customers. For simplification, here we consider w.l.o.g. that for every certain tenant, every service demand can be supported by one slice of the same type. Once a tenant is granted with the requested network slice, it launches its business session to deliver service to the end-users. The duration of a business session, i.e. the lifetime of the corresponding network slice, is a random variable, 
\change{of which the expectation can be estimated by the tenant before issuing the slice request.}

It can take the tenant a one-time cost $u_0$ to issue the service request to the MNO, which is used to issue the request and prepare the end-user service. Besides, this cost may also covers part (or even the whole lump) of the rent for requested network slice, which the MNO may also requires the tenants to prepay as deposit in advance. Additionally, when a request waits in queue, it can consistently generate a periodical waiting cost $u$ for the tenant, which is used to keep the tenant standby for launching the business session. \change{The value of $u_0$ and $u$ shall be carefully tuned by the MNO with respect to its network slice service capacity, selection policies and the tenants' demand, which might be mapped onto an optimal pricing problem, out of scope of this study.}

\change{Upon an admission, a new network slice will be created and granted to the corresponding tenant to support the desired end-user service.} This service is supposed to generate a periodic revenue \textit{Rev} that we assume as known or well predictable by the tenant. Meanwhile, the tenant has to pay a periodical expenditure \textit{Exp} that is composed by the operations cost to maintain the service, and the residential rent for the network slice in case that the rent is not completely prepaid to the MNO in the request-issuing phase. We can assume \change{w.l.o.g}. that the periodical profit $\zeta=\textit{Rev}-\textit{Exp}$ is always positive~--~as the tenant will never issue such a request otherwise.

\subsection{Rational balking \& reneging strategies}\label{sec:rational_strategies}
\change{A request can be characterized by a vector $[n, u_0, u,\zeta,\tau]$, where  $n$ is the slice type, $\tau$ is the expected slice life time upon acceptance, $u_0$, $u$ and $\zeta$ are the business model parameters discussed in Section~\ref{subsec:tenant_business_model}}. Meanwhile, a queue can be characterized by $[l, \lambda,\mu]$ where $l$ is its current queue length, $\lambda$ and $\mu$ are the arrival rate and serving rate, respectively.

When the business demand arises, i.e. request is generated (not issued yet) by the tenant, the expected total profit that this business session can generate is
\begin{equation}
\text{E\{Profit\}}=\zeta\tau.
\end{equation}
Meanwhile, the expected cost of issuing the request and waiting in the queue till acceptance is
\begin{equation}
\text{E\{Cost\}}=u_0+u\text{E}\{w_l\},
\end{equation}
where $w_l$ is the time a request must wait in queue until being accepted, when there are $l-1$ other requests ahead of it.

Assume that the tenants can obtain the a priori knowledge about $w_l$, self-evidently, a rational tenant will issue the request if and only if $\text{E}\{\text{Profit}-\text{Cost}\}\ge0$, which can be described as a binary decision model:
\begin{equation}\label{equ:balking_general}
D_\text{b}=\begin{cases}
1&\zeta\tau-u_0-u\text{E}\{w_l\}\ge0;\\
0&\text{otherwise},
\end{cases}
\end{equation}
where $D_\text{b}=1$ stands for issuing and $D_\text{b}=0$ for balking.

The rational strategy for reneging mostly follows the same principle, but slightly differs from the balking case by concentrating on future cost and neglecting the sunken costs, i.e. the issuing cost and the waiting cost already generated. For an arbitrary request at time $t$, denote with $k(t)$ its current position in queue, and with $w(k)$ its remaining waiting time at position $k$, the tenant is able to rationally choose whether to renege according to:
\begin{equation}\label{equ:reneging_general}
D_\text{r}(t)=\begin{cases}
1&\zeta\tau-u\expectation{w_k(t)}\ge0;\\
0&\text{otherwise},
\end{cases}
\end{equation}
where $D_\text{r}(t)=1$ indicates waiting and $D_\text{r}(t)=0$ for reneging.

\subsection{Rational balking without reneging}\label{subsec:balking_without_reneging}
For simplification, we first ignore reneging, so that
\begin{align}
&\text{E}\{w_l\}=\frac{l}{\mu},\\
&D_\text{b}=\begin{cases}
1&\tau\ge \frac{u_0\mu+ul}{\mu\zeta}\\
0&\text{otherwise}
\end{cases}.
\end{align}

Hence, given certain $\zeta$, $u_0$ and $u$, it yields that $D_\text{b}=D_\text{b}(l,\tau)$ if the tenant is able to observe $l$ before pushing its request into the queue. The balking chance of such a tenant is therefore a function of $l$ under any certain distribution of $\tau$:
\begin{equation}
\begin{split}
b(l)=&\int_{0}^{+\infty}D_\text{b}(l,t)f_{\tau}(t)\text{d}t=\int_{\frac{u_0\mu+ul}{\mu\zeta}}^{+\infty}f_{\tau}(t)\text{d}t\\
=&1-F_{\tau}\left(\frac{u_0\mu+ul}{\mu\zeta}\right)
\end{split}
\end{equation}

Particularly, when
$u_{0}=0$ and $\tau\sim\mathcal{U}(0,\tau_{\max{}})$:
\begin{equation}
b_\text{uni}(l)=\begin{cases}
1-\frac{ul}{\mu\zeta\tau_{\max}}&0\le l\le\frac{\mu \zeta\tau_{\max}}{u};\\
0&\text{otherwise},
\end{cases}
\end{equation}
which is the linear balking model in \cite{ancker1963some1}. Note that there is an implicit limit for the queue length $l_{\max}=\frac{\mu\zeta\tau_{\max}}{u}$, even if no such limit is explicitly set by the MNO.

When $u_{0}=0$ and $f_{\tau}(t)=\frac{1}{(t+1)^2}$ (rational distribution):
\begin{equation}
b_\text{rat}(l)=1-\left.\left(-\frac{1}{t+1}\right)\right\vert_{t=0}^{\frac{ul}{\mu\zeta}}=\frac{\mu\zeta}{ul+\mu\zeta}=\frac{\frac{\mu\zeta}{u}}{l+\frac{\mu\zeta}{u}}.
\end{equation}

When $u_0=0$ and $\tau\sim\text{Par}(1,1)$:
\begin{equation}
b_\text{par}(l)=\begin{cases}
1&l=0;\\
1-\left.\left(-\frac{1}{t}\right)\right\vert_{t=1}^{\frac{ul}{\mu\zeta}}=\frac{\frac{\mu\zeta}{u}}{l}&\text{otherwise},
\end{cases}
\end{equation}
which is the hyperbolic balking model in \cite{ancker1963some2} with the factor of patience $\beta=\frac{\mu\zeta}{u}$. Note that this model assumes every slice remains active for at least an unit time period.

When 
$u_0=0$ and $\tau\sim\text{Exp}(\eta)$:
\begin{equation}
b_\text{exp}(l)=1-\left.\left(-e^{-\eta t}\right)\right\vert_{t=0}^{\frac{ul}{\mu\zeta}}=e^{-\frac{\eta u}{\zeta}\frac{l}{\mu}},
\end{equation}
which is the exponential balking model in \cite{shortle2018fundamentals} with the exponent of impatience $\beta=\frac{\eta u}{\zeta}>0$. 

\subsection{Rational reneging} \label{subsec:reneging}
Now we consider the reneging behavior. As we will derive below, the decision of reneging highly relies on the tenant's knowledge of the queue. So we discuss this problem separately in different cases.

\subsubsection{Full knowledge}
First, we assume that every tenant is not only able to observe the position $k$ of its request in the queue in real time, but also informed by the MNO about $\expectation{w_k}$.  In this case,
%
Eq.~\eqref{equ:reneging_general} can be directly applied to determine the maximal waiting time $t_{\max}$:
\begin{equation}
\zeta\tau -u\text{E}\{w_{k(t_{\max})}\}=0\Rightarrow
\text{E}\{w_{k}\}=\sum\nolimits_{i=0}^{k-1}\frac{1}{\mu+\sum\nolimits_{j=0}^{i}\omega_j},\label{equ:tmax_general}
\end{equation}
where $\omega_i\ge0$ is the reneging rate at the queue position $i$ for $i>0$ and $\omega_0=0$. Therefore, this case has an equivalent form where the MNO informs the tenant that issues the $k^\text{th}$ request in queue about $\mu$ and $\omega_i$ for all $i<k$. The values of $\mu$ and $\omega_i$ converge to stable levels in long term when the business scenario remains consistent, therefore we consider them here as constants that are known by the MNO.

\subsubsection{Knowledge of serving rate} 
In this case, we assume that every tenant is informed by the MNO about $\mu$, and keeps observing the position $k$ of its request in the queue, but has no knowledge about $\omega_i$. As a tenant generally lacks knowledge of requests issued by other tenants, i.e. the statistics of their parameters $[u_0,u,p,\tau]$. So no tenant is able to estimate the values of $\omega_i$ in this case, which disables the estimation of $\expectation{w_{k}}$ according to Eq.~\eqref{equ:tmax_general}. However, knowing that $\omega_i\ge0$ for all $i$, the tenants can make conservative estimations based on
\begin{equation}\label{equ:conservative_known_mu}
\expectation{w_{k}}=\sum\nolimits_{i=0}^{k-1}\frac{1}{\mu+\sum\nolimits_{j=0}^{i}\omega_j}\ge\frac{k}{\mu},
\end{equation}
and therefore Eq.~\eqref{equ:reneging_general} becomes
\begin{equation}\label{equ:reneging_serving_knlowledge}
D_\text{r}(t)=\begin{cases}
1&k(t)\le \frac{\mu\zeta\tau}{u};\\
0&\text{otherwise}.
\end{cases}
\end{equation}

\subsubsection{Knowledge of position}\label{subsubsec:knlg_of_pos}
In this case, every tenant is able to track its request's current position $k$ in queue, but has no a priori knowledge about $\mu$. Thus, the tenant has to estimate $\mu$ through an online learning process while waiting in queue:
\begin{equation}
\hat{\mu}(k)=\frac{l-k}{T_k},
\end{equation}
where $T_k$ is the time that the request took to arrive the $k^\text{th}$ position since its entrance to the queue.
Thus, Eqs.~\eqref{equ:conservative_known_mu} and \eqref{equ:reneging_serving_knlowledge} become respectively
\begin{align}
&\expectation{\hat{w}_{k}}=\sum\nolimits_{i=0}^{k-1}\frac{1}{\hat{\mu}(k)+\sum\nolimits_{j=0}^{i}\omega_j}\ge\frac{k}{\hat{\mu}(k)}=\frac{kT_k}{l-k},\label{equ:estimation_mu}\\
&D_\text{r}(t)=\begin{cases}
1&k(t)\le \frac{l\zeta\tau}{uT_{k(t)}+\zeta\tau};\\
0&\text{otherwise}.
\end{cases}
\end{align}

It has to be noted here that the estimation in Eq.~\eqref{equ:estimation_mu} is only valid when $k<l$, and the estimation error $\epsilon_\mu^2$ decreases as $k$ increases. Therefore a threshold $\Delta K$ should be set whereas $\hat{\mu}(t)$ is estimated only when $l-k\ge \Delta K$. 
In summary:
\begin{equation}
D_r(t)=\begin{cases}
1&k(t)<l-\Delta K,\\
1& l-\Delta K\le k(t)\le \frac{l(\zeta\tau}{uT_{k(t)}+\zeta\tau},\\
0&\text{otherwise}.
\end{cases}
\end{equation}

\subsubsection{Knowledge of average waiting time}
In this case, all tenants are only informed about the average waiting time $\overline{w}$ since joining the queue till being served, in which the waiting time requests that balk and renege do not count. Meanwhile, the current position in queue $k$ is unobservable for a request unless $k=0$ (i.e. when the request gets served).  

In this case, a request can only roughly consider all other requests ahead of it in queue as a batch. Since the service to every single request is a Poisson event, we approximately consider the complete service to this batch (of unknown length) as a Poisson event with arriving rate of ${1}/{\overline{w}}$. Thus, the reneging decision can be made as
\begin{equation}\label{equ:rng_avg_wait_time_knlg}
D_\text{r}(t)=\begin{cases}
1&\zeta\tau-u\overline{w}\ge0;\\
0&\text{otherwise}.
\end{cases}
\end{equation}
Note that Eq.~\eqref{equ:rng_avg_wait_time_knlg} is independent of $t$ or $k$ so that it always returns the same decision.

\subsubsection{Blindness}\label{subsubsec:blindness}
If the tenant possesses neither the position $k$ of its waiting request in the queue, nor any knowledge about the dynamics of queue, it can only make a blind reneging, where a maximal waiting time is predetermined at the queue entrance. A straightforward solution is to set a maximal cost proportional to the total profit that can be generated by the requested slice upon admission:
$u_{\max{}}=u_0+ut_{\max{}}=\ni\zeta\tau$,
where $\ni\ge 0$ is the factor of risking that indicates the tenant's intension of waiting in the queue. 
This yields that 
\begin{equation}
D_\text{r}(t)=\begin{cases}
1&t<\frac{\ni\zeta\tau-u_0}{u},\\
0&\text{otherwise}.
\end{cases}
\end{equation}
It is worth to note that when $u_0=0$ and $\tau\sim\text{Exp}(\eta)$, the blind reneging model becomes the classic reneging model \cite{ancker1963some2} where the maximal waiting time is exponentially distributed. Furthermore, when $\ni\to+\infty$ the tenants will never renege and therefore become \emph{patient}.

\subsection{Balking with renaging}\label{subsec:balking_with_reneging}
When tenants are able to renege on their requests and the issuing cost $u_0=0$, the balking behavior can be considered as a special case of reneging at the queue entrance ($t=0, k=l$). Clearly, this implies to disable balking in the aforementioned cases \ref{subsubsec:knlg_of_pos} and \ref{subsubsec:blindness}  where no a priori estimation of $\text{E}\{w\}$ is available for tenants.

\section{Slice admission strategy optimization}\label{sect:optimization}
In slice admission control, there are various performance metrics that may include:\change{ the overall network utility, the admission rate, and the average request waiting time, as discussed below}.

The network utility rate of a slice can be differently defined, such as the periodical payment $u$ that the MNO receives from the tenant, or the generated network throughput, etc. It is common to consider the utility rate of a slice as determined by the slice type, and the overall network utility rate at any time instant $t$ as the sum of utility rates of all active slices:
\begin{equation}\label{equ:instant_net_utility_rate}
u_\Sigma(t)=\sum\nolimits_{n=1}^{N}s_n(t)u_n,
\end{equation}
where $s_n(t)$ is the number of type-$n$ active slices at time $t$, and $u_n$ is the utility rate of every type-$n$ slice. In long term, the average overall network utility rate can be estimated from the acceptance and releasing rates of different slice types:
\begin{equation}\label{equ:mean_net_utility_rate}
\overline{u}_\Sigma=\sum\nolimits_{n=1}^{N}\frac{\mu_n u_n }{\eta_n}.
\end{equation}

The average waiting time of all requests in queues is
\begin{equation}
\overline{W}_\text{q}=\frac{\sum\nolimits_{n=1}^{N}\overline{W}_{\text{q},n}L_n}{\sum\nolimits_{n=1}^{N}L_n}.
\end{equation}
The overall admission rate is the following
\begin{equation}
\overline{P}(A)=\frac{\sum\nolimits_{n=1}^N\lambda_nP(A_n)}{\sum\nolimits_{n=1}^N\lambda_n}.
\end{equation}

All three criteria are determined by the request behavior parameters $\alpha_n,\beta_n,\lambda_n$ and the acceptance rate $\mu_n$. Given a certain combination of $[\alpha_n,\beta_n,\lambda_n,\eta_n]$,
$\mu_n$ is uniquely determined by the MNO's strategy, i.e. by the preference matrix $\mathbf{\Phi}$. Hence, with consistent behaviors of request arrival and slice releasing, we can optimize either of them by selecting the best $\mathbf{\Phi}$.

A major challenge for analysis exists in the complex relation between the acceptance rates $[\mu_1,\mu_2,\dots,\mu_N]$  and the strategy $\mathbf{\Phi}$, as $\mathbf{\Phi}$ does not directly imply the MNO's action or statistics, but only its preference. 

Nevertheless, if the steady-state probability of queue lengths $p_n(l)$, as defined in Eq.~\eqref{equ:steady_queue_state}, is known or measurable for all $n\in\mathcal{N}$, we can estimate $\mu_n$ for all $n$ with respect to $\mathbf{\Phi}$ and the initial state $\mathbf{s}_\text{init}$ as follows.

First, define a bijection $\mathbb{S}\leftrightarrow\{1,2,\dots,\vert\mathbb{S}\vert\}$ as $J=J_\mathbb{S}(\mathbf{s})$ where $J_\mathbb{S}(\mathbf{s})=I_\mathbb{A}(\mathbf{s})$ for all $\mathbf{s}\in\mathbb{A}$. Then extend the definitions in Eqs. (\ref{equ:slice_incremental_vector}), (\ref{equ:phi_matrix}) and (\ref{equ:steady_queue_state}) with
\begin{align}
&\Delta\mathbf{s}_0=\underbrace{[0,0,\dots,0]}_N,\\
&\tilde\phi_{i,j}=\begin{cases}
0&j>\vert\mathbb{A}\vert\\
\phi_{i,j}&j\le\vert\mathbb{A}\vert
\end{cases},\forall i\in\{1,2,\dots,N+1\},\\
&p_{0}(0)=0,
\end{align} 
respectively. The probability of state transition from $\mathbf{s}\in\mathbb{S}$ to $\mathbf{s}+\Delta\mathbf{s}$ can be then calculated as
\begin{equation}
\text{Prob}(\mathbf{s}\to\mathbf{s}+\Delta\mathbf{s}_n)=\prod_{k=1}^{n-1}p_{\tilde\phi_{k,J}}(0)(1-p_{\tilde\phi_{n,J}}(0)).
\end{equation}

Thus, when the initial state $\mathbf{s}_\text{init}$ is known, we can obtain the long-term probability distribution of system state $\mathbf{s}$ as
\begin{equation}
\text{Prob}(\mathbf{s}_j~\vert~\mathbf{s}_\text{init}=\mathbf{s}_i)=\lim\limits_{K\to\infty}\frac{1}{K}\sum\nolimits_{k=0}^{K}[\mathbf{\Psi}^k]_{i,j},
\end{equation}
where $\mathbf{\Psi}$ is the transition matrix where $\Psi_{i,j}=\text{Prob}(\mathbf{s}_i\to\mathbf{s}_j)$:
\begin{equation}
\mathbf{\Psi}=\begin{bmatrix}
\Psi_{1,1} &\Psi_{1,2} & \dots & \Psi_{1,\vert\mathbb{S}\vert}\\
\Psi_{2,1} &\Psi_{2,2} & \dots & \Psi_{2,\vert\mathbb{S}\vert}\\
\vdots&\vdots&\ddots&\vdots\\
\Psi_{\vert\mathbb{S}\vert,1} &\Psi_{\vert\mathbb{S}\vert,2} & \dots & \Psi_{\vert\mathbb{S}\vert,\vert\mathbb{S}\vert}\\
\end{bmatrix}.
\end{equation}

More generally, if not the exact value but the  probability distribution of  the initial state is available as $P_{\text{init}}=[p_{\text{init}}(\mathbf{s}_1),p_{\text{init}}(\mathbf{s}_2),\dots,p_{\text{init}}(\mathbf{s}_{\vert\mathbb{S}\vert})]$, the long-term probability distribution $\mathbf{s}$ is the following
\begin{equation}
\text{Prob}(\mathbf{s}_j~\vert~P_\text{init})=\lim\limits_{K\to\infty}\frac{1}{K}\sum\nolimits_{k=0}^{K}\sum\nolimits_{i=1}^{\vert\mathbb{S}\vert}p_{\text{init}}(\mathbf{s}_j)[\mathbf{\Psi}^k]_{i,j}.
\end{equation}

We can obtain the expected active slice number $\overline{s}_n$ of every slice type $n$ as a function of $\mathbf{\Psi}$ and thus, as a function of $\mathbf{\Phi}$. Now, recalling Eqs.~(\ref{equ:instant_net_utility_rate}--\ref{equ:mean_net_utility_rate}) it yields that
\begin{equation}
\overline{s}_n=\frac{\mu_n}{\eta_n},
\end{equation}
and then we can write the following
\begin{equation}\label{equ:cost_function}
\begin{split}
\mu_n&=\frac{\overline{s}_n}{\eta_n}=\frac{\sum\nolimits_{\mathbf{s}\in\mathbb{S}}\text{Prob}(\mathbf{s}~\vert~P_\text{init})s_n}{\eta_n}\\&
=\frac{1}{\eta_n}\sum\nolimits_{\mathbf{s}\in\mathbb{S}}\lim\limits_{K\to\infty}\frac{1}{K}\sum\nolimits_{k=0}^{K}\sum\nolimits_{i=1}^{\vert\mathbb{S}\vert}p_{\text{init}}(\mathbf{s}_j)[\mathbf{\Psi}^k]_{i,j}.
\end{split}
\end{equation}

Based on this analytical expression, we are able to optimize $[\mu_1,\mu_2,\dots,\mu_n]$ with respect to $\mathbf{\Phi}$. However, it is evident that Eq.~\eqref{equ:cost_function} is non-convex w.r.t. $\mathbf{\Phi}$, which prohibits analytical solution of the global optimum. On the other hand, the overall domain size of $\mathbf{\Phi}$ is $2^{(N+1)\vert\mathbb{A}\vert}$, which can assume unaffordable high values for any realistic dimension of $\vert\mathbb{A}\vert$ in practical networks, making the exhaustive search impossible. \change{This is an integer linear programming (ILP) problem that is proven to be NP-Hard. Advanced machine learning and heuristic search methods might be explored to solve it at an affordable computational load. Due to space constraints this is left as future work.}

%
%

\section{Numerical simulations}
\label{sect:perf_eval}

\subsection{Simulating the decisions of impatient tenants}\label{subsec:num_eva}
\subsubsection{Simulation setup}\label{subsec:sim_setup}
In the numerical simulations we define a simplified scenario, where the MNO holds a normalized two-dimensional ($M=2$) resource pool $\mathbf{r}=[1,1]$ and $N=2$ different slice types are defined and specified with the parameters listed in Table \ref{tab:slice_types}. Note that we assume the lifetime of every type-$n$ slice is an exponentially distributed random variable $\tau_n\sim\text{Exp}(\eta_n)$, where $1/\eta_n$ is the average lifetime of type-$n$ slices.
\begin{table}[!htbp]
	\centering
	\caption{\change{Specifications of the defined slice types (all parameters normalized to dimensionless quantity)}.}
	\label{tab:slice_types}
	\small\begin{tabular}{c|c|c|c|c|c|c}
		\toprule[1.2px]
		\textbf{Slice type ($n$)} &$\mathbf{u}_n$ & $\lambda_n$ & $1/\eta_n$ & $u_{0,n}$ & $u_n$ & $\zeta_n$\\\hline
		1 & $[0.01,0.05]$ & 6 & 5 & 0 & 1 & 8\\\hline
		2 & $[0.05,0.01]$ & 10 & 3 & 0 & 1.5 & 12\\
		\bottomrule[1.2px]
	\end{tabular}
\end{table}

Under these specifications, the admissibility region $\mathbb{A}$ is composed of $341$ different values of $\mathbf{s}$. For the sake of simplicity, we do not consider the option of ``reserve'' in the MNO's slicing strategy: in this way for every state $\mathbf{s}$ only two different preference vectors, i.e. $\varphi_1=[1,2,0]$ and $\varphi_2=[2,1,0]$, are available. Therefore, there are in total $2^{341}$ different slicing strategies applicable for the MNO. We randomly select $1000$ from these valid slicing strategies, and with every MNO slicing strategy, we evaluate the rational balking/reneging strategies of impatient tenants with different information available. For every individual evaluation, we simulate the arrivals of tenant requests and the MNO's operations for $1000$ periods. To mitigate divergences of queue lengths, especially for the case of patient tenants, an upper bound of length is set to $100$ for every queue.

\subsubsection{Evaluation results}\label{subsec:eva_results}
\begin{table*}
	\centering
	\caption{Tenant profits in 1000 operation periods under different balking/reneging strategies, ``Patience'' is the benchmark strategy where no balking or reneging takes place.}
	\label{tab:sim_results}
	\small\begin{tabular}{ll|c|c|c|c|c|c}
		\toprule[1.2px]
		\multirow{2}{*}{\textbf{Case}}&&\multicolumn{2}{c|}{\textbf{Total profit ($\times10^3$)}}&\multicolumn{2}{c|}{\textbf{Mean profit}}&\multicolumn{2}{c}{\textbf{Profiting chance}}\\\cline{3-8}
		&&Type 1&Type 2&Type 1&Type 2&Type 1&Type 2\\\hline
		\multicolumn{2}{l|}{Patience}&36.06&16.88&10.74&2.86&49.21\%&40.48\%\\\hline
		\multirow{3}{*}{Blindness}&\multicolumn{1}{|l|}{$\ni=1$}&33.62&24.08&7.93&3.54&46.27\%&43.47\%\\\cline{2-8}
		&\multicolumn{1}{|l|}{$\ni=0.1$}&110.89&139.23&18.45&13.90&25.42\%&22.93\%\\\cline{2-8}
		&\multicolumn{1}{|l|}{$\ni=0.01$}&129.45&153.98&21.53&15.40&46.29\%&43.39\%\\\hline
		\multicolumn{2}{l|}{Knowledge of position ($\Delta K=2$)}&36.03&16.86&10.72&2.79&49.23\%&40.47\%\\\hline
		\multicolumn{2}{l|}{Knowledge of average waiting time}&50.88&60.11&32.97&25.36&85.65\%&79.18\%\\\hline
		\multicolumn{2}{l|}{Knowledge of serving rate}&93.12&100.07&57.17&43.25&94.80\%&83.59\%\\\hline
		\multicolumn{2}{l|}{Full knowledge}&92.68&101.53&57.66&44.34&95.57\%&83.87\%\\
		\bottomrule[1.2px]
	\end{tabular}
\end{table*}

During the simulation, we track the end-profit of every issued slice request defined as follows
\begin{equation}
\zeta_\text{e}=\begin{cases}
\zeta\tau-u_0-uw&\text{accepted};\\
-u_0-uw&\text{reneged},
\end{cases}
\end{equation}
where $w$ is the total waiting time from queue entrance to admission/reneging. Then we evaluate the balking/reneging strategies of tenants with three different metrics:
\begin{itemize}
	\item \emph{Total profit}: the sum of end-profits obtained by all issued slice requests;
	\item \emph{Mean profit}: the average end-profit obtained by all issued slice requests;
	\item \emph{Profiting chance}: the ratio of slice requests that lead to positive end-profits in all issued requests.
\end{itemize}
The simulation results are listed in Table~\ref{tab:sim_results}. 

It can be easily observed from the results that, given the knowledge about position of its request in queue and the queue's serving rate, a tenant has a high chance to make correct decisions of balking and reneging. Thereby it is able to mitigate most losses caused by excessive waiting in case of request congestion, and thus obtain a positive profit.

The information about reneging rates provides a further improvement in addition, but only by an insignificant degree. One reason of this phenomena could be that, after a rational balking with sufficient knowledge, the reneging rate of requests generally remains limited, and therefore it exhibits a little impact on the waiting time in queue.

In contrast, when provided with only insufficient information, tenants are likely to benefit less from their impatience. An impatient tenant knowing only the mean waiting time in queue can reasonably avoid most extreme long waitings by balking, and therefore has more chance to achieve a positive profit in comparison to patient tenants, yet significantly lower than the tenants with full knowledge. The knowledge about current position of request in queue alone fail to assist tenants with their decisions, resulting to a similar performance as that of the patient tenants -- which is the bound provided by the artificial queue truncation. Unwise decisions are also made by the blind tenants that renege after predetermined waiting time, whose performance strongly depends on the patience factor $\ni$, and in the worse case even outperformed in profiting chance by patient tenants.

In summary, network slice tenants need information about queue dynamics from the MNO---at least the minimum information to enable balking---so that they can benefit from impatience in case of slice requests congestion.

\subsubsection{Distribution of reneging time}\label{subsec:dist_rng_tmie}
In Section~\ref{subsec:balking_without_reneging}, we have analytically proven the applicability of various classical models of balking statistics in the slice admission control scenario upon different distributions of the slice lifetime $\tau$. The distribution of reneging time in SAC, however, is relatively challenging to derive in such way. 

\begin{figure}[!htbp]
	\centering
	\begin{subfigure}[c]{\flexwidth}
		\centering
		\includegraphics[width=.9\textwidth]{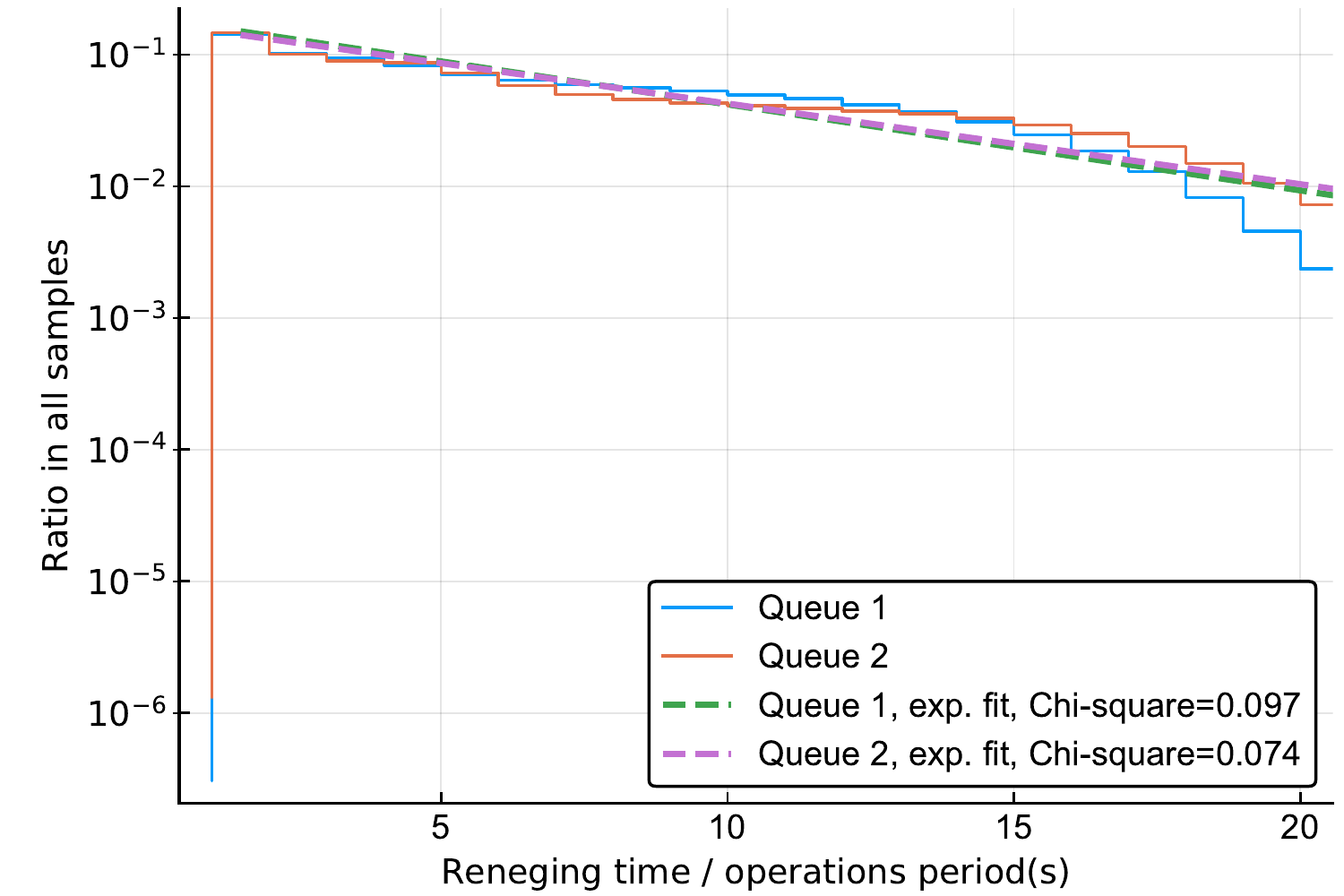}
		\subcaption{Random admission strategy}
	\end{subfigure}
	\hfill
	\begin{subfigure}[c]{\flexwidth}
		\centering
		\includegraphics[width=.9\textwidth]{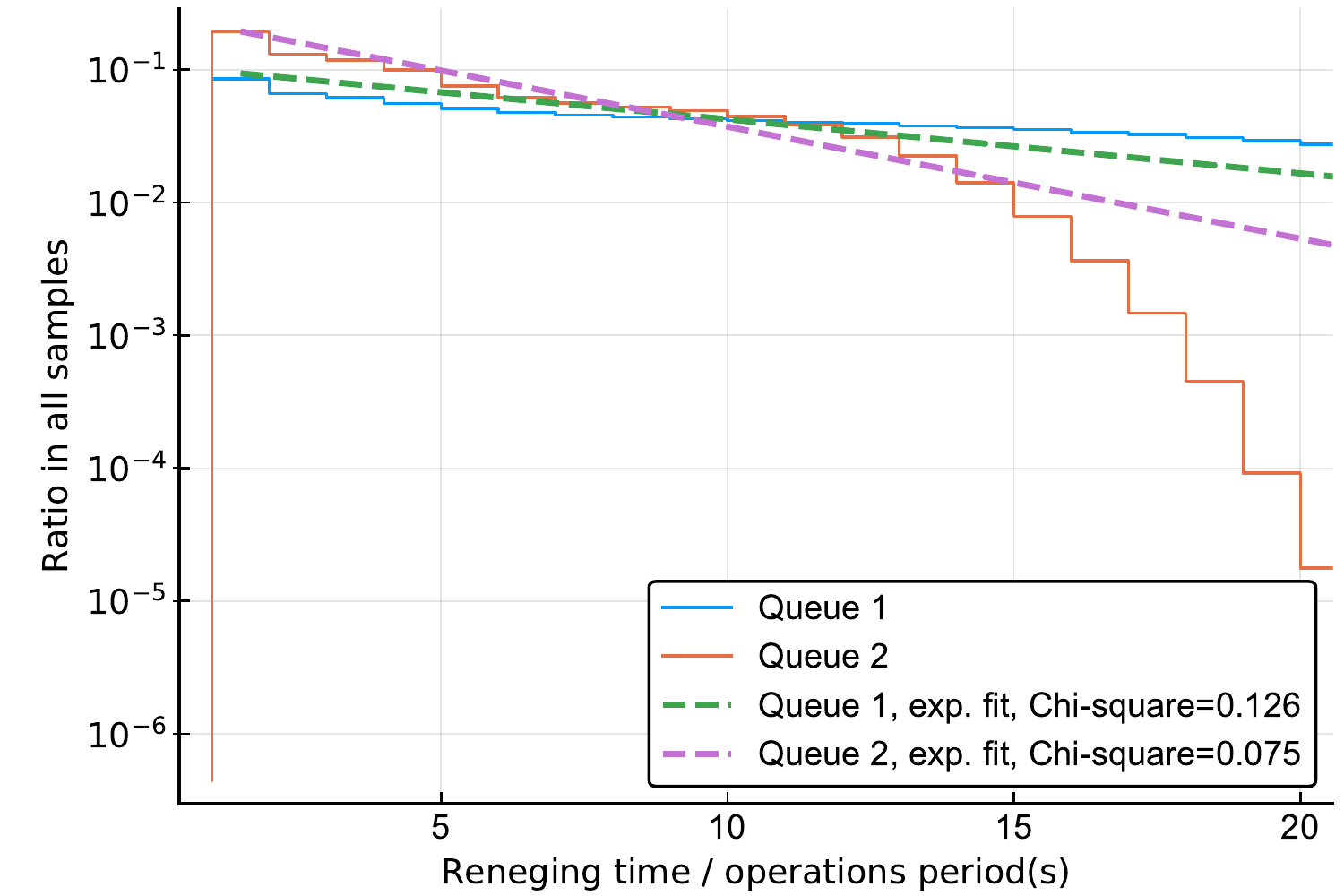}
		\subcaption{Fixed admission strategy (prefer type 2)}
	\end{subfigure}
	\caption{\change{The distribution of tenant requests' reneging time in multi-queue slice admission control with fitting results.}}
	\label{fig:rng_time_distribution}
\end{figure}

To evaluate the applicability of existing reneging models, we execute additional numerical simulations. The environment is configured to the same specifications listed in Table~\ref{tab:slice_types}, and the tenants possess full knowledge of the queuing system. First we randomly generate $1000$ different admission strategies, for every strategy we carry out $25$ rounds of Monte-Carlo test, in each round the MNO operates $40$ periods. Then we fix the MNO to a static admission strategy that $\varphi=[2,1,0]$ for all $\mathbf{s}\in\mathbb{A}$, and repeat this test $1000$  times, also with $25$ rounds per time and $40$ operations periods per round. We observe the waiting time of all reneged requests and illustrate the obtained results in Fig.~\ref{fig:rng_time_distribution}. It can be observed that the exponential distribution generally provides a satisfactory fit to the reneging time in most cases, which supports applying the classical model proposed in \cite{ancker1963some2} to simplify queue models from the MNO's perspective. However, it shall be remarked that in case of strong congestion where a queue can be extremely long, e.g. Queue 1 in the fixed strategy test, the reneging time may become fat-tail distributed and no more exponential.

\subsubsection{Impatience model selection}\label{subsec:model_choice}
Certainly, when fed with the knowledge of the current active queues, tenants may be more encouraged to balk or renege from densely congested queues of slice requests, which in turn leads to a decreased number of awaiting slice requests. Nevertheless, it should be noted that the phenomena of balking and reneging are only significant when the queues are considerably long. In this case, the MNO's resources are already sufficiently utilized, and the utilization rate is hardly impacted by the impatience of tenants. On the other hand, if there is a lack of information about the queues, as demonstrated in Section~\ref{subsec:eva_results}, tenants can suffer from high probability of business loss. This will, self-evidently, suppress the tenants' interest for the MNO's slice service on long-term windows, leading to a consistent loss of customers from the MNO's perspective. In summary, we can argue that it is a \emph{win-win option for the MNO to share full knowledge of the queues}, or at least the request's current position in queue and the serving rate of queue, to every awaiting tenant. In such context, we assert it to be reasonable and rational to apply the models of exponential balking and reneging, as we did in Section~\ref{subsec:performances_with_balking_and_reneging}.

\subsection{Simulating the heterogeneous multi-queue slice controller}
To carry out simulations in a consistently specified environment, we consider the same scenario as defined in Table~\ref{tab:slice_types}.

\subsubsection{Verification of geometric IAT distribution}\label{subsec:geometric_iat}
\begin{figure}[!htbp]
	\centering
	\begin{subfigure}{\flexwidth}
		\centering
		\includegraphics[width=.9\columnwidth]{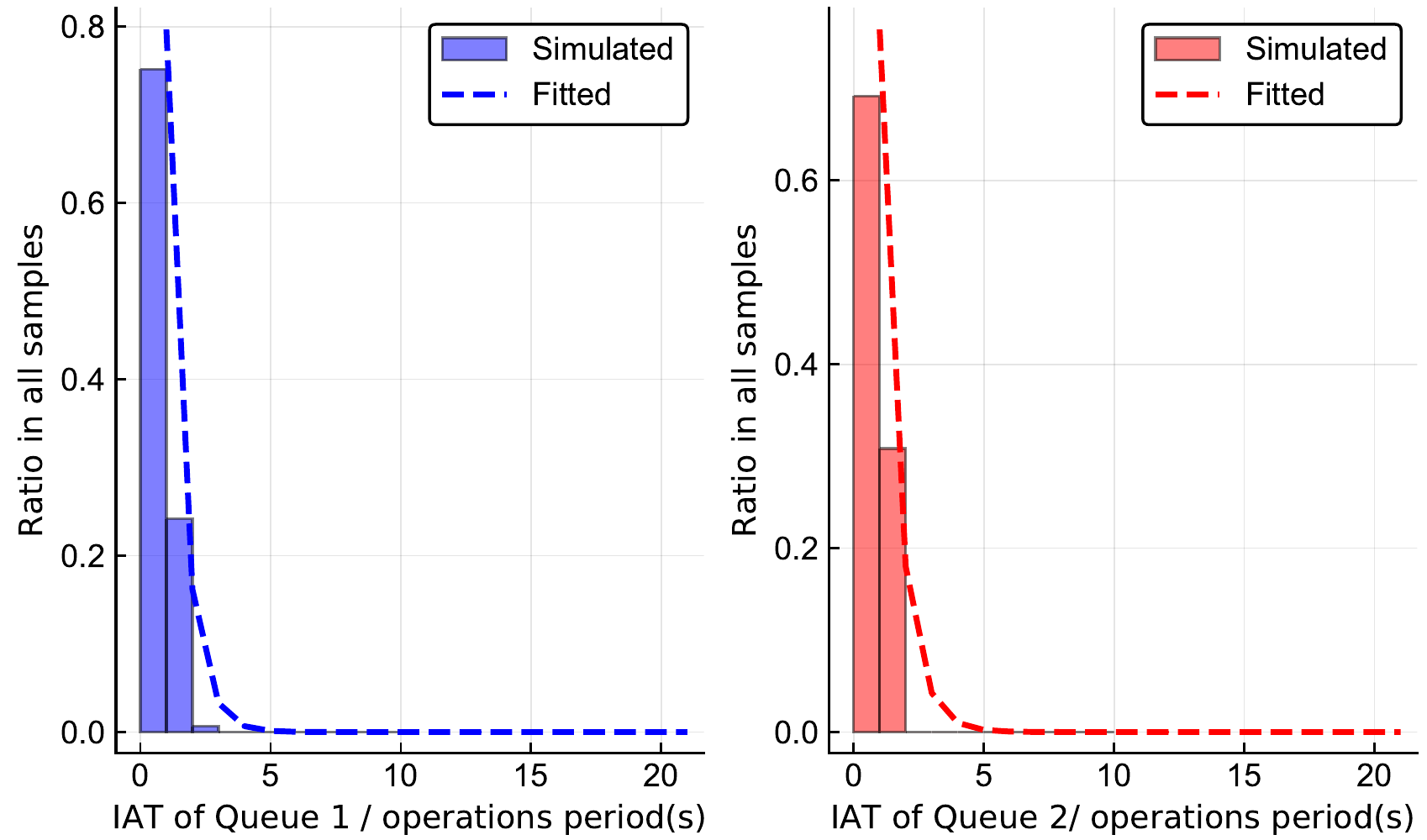}
		\caption{The distribution of inter-acceptance time in two different queues under a random strategy, fitted as geometric distribution.}
		\label{subfig:geometric_acceptance_sample}
	\end{subfigure}
	\begin{subfigure}{\flexwidth}
		\centering
		\includegraphics[width=.9\columnwidth]{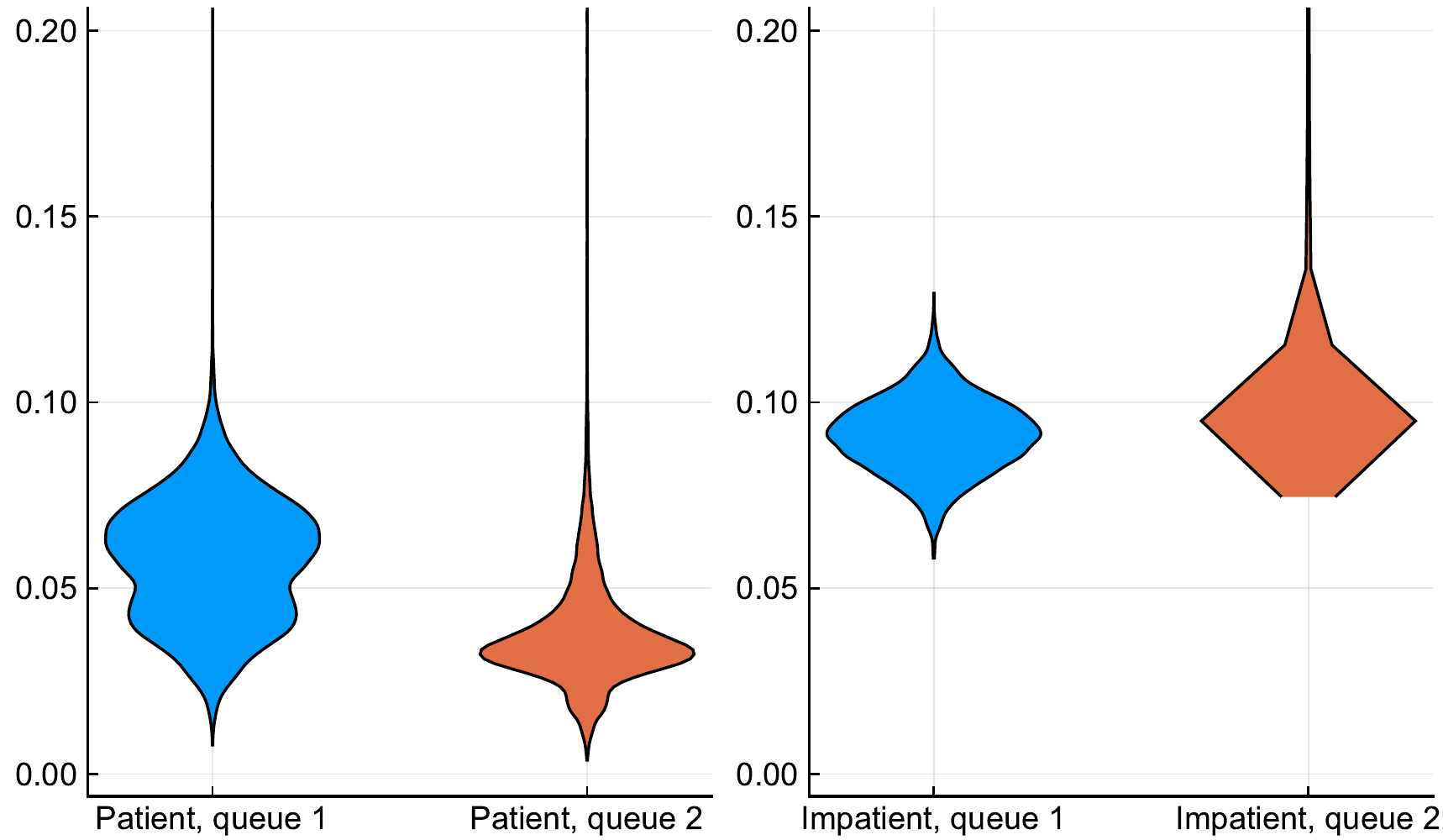}
		\caption{The KLD of fitting IAT with geometric distribution, 1000 random strategies tested, 25 Monte Carlo tests for each. Fitting success rate is $99.948\%$ for patient tenants and $20.568\%$ for impatient ones.}
		\label{subfig:geometric_acceptance_KLD}
	\end{subfigure}
	\caption{\change{The IAT of individual queue under an arbitrary strategy is geometrically distributed \emph{iff} tenants are patient.}}
	\label{fig:geometric_acceptance}
\end{figure}

In case of patient tenants, \change{Remark} \ref{remark:geom_iat} can also be verified through numerical simulations. We consider all tenants as patient, and set an upper bound of queue length to 100 for all queues to mitigate queue divergence. Then we randomly generate $1000$ slicing strategies, in which the resource reservation ($n=0$) is always assigned with the least preference. For each strategy, $25$ rounds of Monte-Carlo tests are executed. In each testing round, an MNO with a 2-queue slice admission controller is initialized to a random but fully resource-utilized state, and then operates under the consistent strategy for $40$ operations periods. Then we investigate the distribution of inter-acceptance time (IAT) for each queue, and attempt to fit the measurements with geometric distributions, which is the discrete-time version of exponential distribution. Out of all the $25~000$ tests, $99.948\%$ IAT records were successfully fitted by a Maximum-Likelihood-Estimator (MLE). A sample result is shown in Fig.~\ref{subfig:geometric_acceptance_sample}, where a good fitting performance can be observed. 

To evaluate the fitting for every strategy we use the Kullback-Leibler divergence (KLD)~\cite{kullback1997information}:
\begin{equation}
D_\text{KL}(P_\text{IAT}~\vert~\text{Geom.})=\sum\nolimits_{k=0}^{\infty}p_\text{IAT}(k)\log\frac{p_\text{IAT}(k)}{(1-\hat{p})^k\hat{p}},
\end{equation}
where $p_\text{IAT}(k)$ is the empirical probability mess function (PMF) of the measured IAT, and $(1-\hat{p})^k\hat{p}$ is the geometric PMF with fitted parameter $\hat{p}$. KLD is an indicator of fitness between two distributions, which equals $0$ for two identical distributions and approaches towards $+\infty$ for two completely irrelevant ones. The KLD distribution over all $25~000$ tests is depicted in the left part of Fig.~\ref{subfig:geometric_acceptance_KLD}, which shows a satisfactory fitness for both queues (slice types).

Furthermore, to verify the impact of impatient tenants' behavior, we grant all tenants with full knowledge about the queues to activate balking and reneging, and then repeat the aforementioned simulation procedure. Only $20.568\%$ of the measured IAT tracks can be successfully fitted with the geometric distribution this time (on the rest measurements, the MLE fails to converge). The KLD distribution of successfully fitted IAT tracks is illustrated in the right part of Fig.~\ref{subfig:geometric_acceptance_KLD}. Compared to the case of patient tenants, we can observe a significant increase of KLD here, confirming our assertion that the behaviors of balking and reneging will remove the Markovian feature of the system. Remark that when the balking and reneging rates are low, such impact can be slight enough to be neglected. 

\subsubsection{Evaluation of the proposed controller}
To verify the effectiveness and potential in optimization of the proposed multi-queue slice admission controlling mechanism, we generate $10~000$ random strategies, and measure all three above-mentioned performances metrics $\overline{u}_\Sigma$, $\overline{W}_\text{q}$ and $\overline{P}(A)$ for every strategy in both reference scenarios $1$ and $2$. Similar to the last tests, every strategy is evaluated through a $25$-round Monte-Carlo test where each round begins with a random initial state and lasts $40$ operations periods. Impatient tenants are considered.

To provide benchmarks, we test the controller with two specific ``na\"ive'' strategies: \emph{Prefer type 1}: the preference vector is $[1, 2, 0]$ at all system states; \emph{Prefer type 2}: the preference vector is $[2, 1, 0]$ at all system states.
Moreover, we implement and test a simple ``greedy'' single-queue slice admission controller that always accepts the first request in its queue regardless of type, as long as the resource pool supports.

\begin{figure}[!htbp]
	\centering
	\includegraphics[width=0.8\flexwidth]{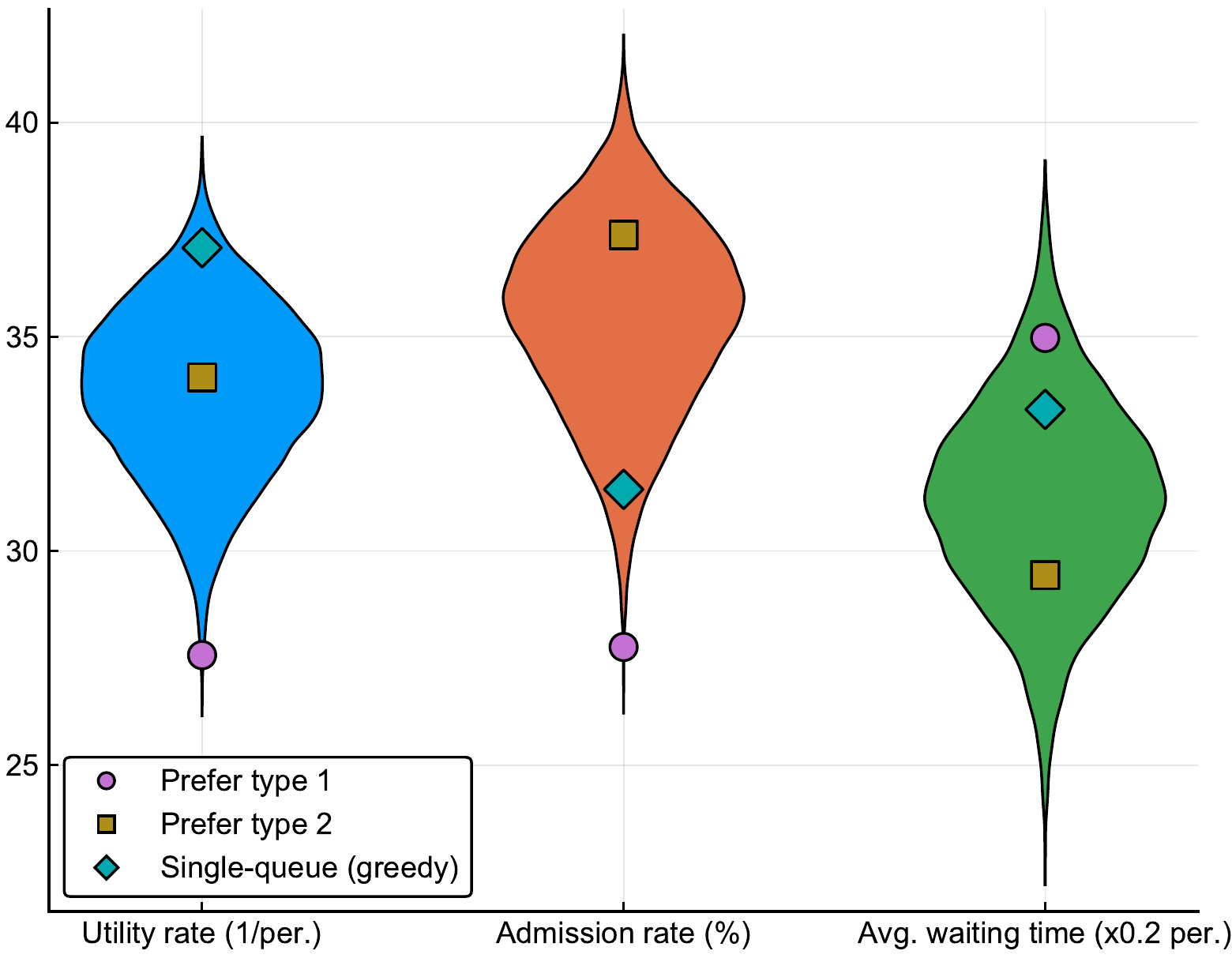}
	\caption{Performance distribution of the proposed multi-queue slice admission controller with $10~000$ random strategies, in comparison to selected benchmarks.}
	\label{fig:effectiveness}
\end{figure}

The results are illustrated in Fig.~\ref{fig:effectiveness}. It can be observed that the multi-queuing controller, when specified with an appropriate strategy, outperforms the greedy single-queue solution, especially when under dense demand and queue are congestions. Note that, however, the performance highly relies on the strategy selection, leading to a critical necessity of strategy optimization.

\section{Further discussion}\label{sect:discussions}
In practical wireless networks, both the dynamics of resource availability (e.g. channel fading) and the resource elasticity of active slices must be taken into account. The model in this paper is an approximation with a static resource pool $\mathbf{r}$ and rigid slices, which holds in long-term with appropriate dynamic scheduling to multiplex slices. Note that such a slice multiplexing implicitly enables slice \emph{overbooking} with a risk to break SLAs~\cite{zanzi2018ovens,ovnes2018conext}. The challenge of balancing the multiplexing gain and the overbooking risk in heterogeneous multi-queue admission control settings deserves future study.

It shall also be noticed that the assumptions of Poisson arrivals/releases may not hold in some practical service scenarios. In this case, the queues are not $M/M/1$ systems and cannot be considered as continuous-time Markov systems. Nevertheless, as pointed out in \cite{shortle2018fundamentals},
many such continuous-time non-Markov processes can be easily transformed into discrete-time Markov chains by observing only the state transitions.
Therefore, the analyses given above also apply to most scenarios with non-Poisson request arrivals/releases.

\vspace{+2mm}
\section{Related work}
\label{sect:rel_work}

We summarize the main research effort in the literature on different topics, such as Slice-as-a-Service, queuing theory for cloud services and network slicing admission control. 

An overview on multi-tenancy service and 5G network slicing is given in~\cite{samdanis2016network} from perspectives of architecture and standardization, introducing the novel concept of \emph{network slice broker}, which executes the admission control. Different attempts have been made in~\cite{sciancalepore2017slice,bega_TMC2019,addad_TMC2019} and~\cite{sciancalepore2019mobile} to demonstrate how the admission control can benefit the overall network resource utilization. In~\cite{Wen_TC2019}, a robust network slicing mechanism by addressing the slice recovery and reconfiguration in a unified framework
Additionally, in~\cite{han2019utility} a multi-queuing system for heterogeneous tenant requests is modelled to derive statistical behavior models showing how this can be approximated to a Markovian system. However, none of the above-mentioned works have addressed the option of allowing infrastructure provider to share information with upcoming tenants so as to improve the overall system performance while, at the same time, formulating a network slice admission optimization problem based on a novel utility model. 

While we have considered network slicing in a generic and abstracted view, which is generally applicable in both radio access network (RAN) and core network (CN) domains, recently there has been a dense specific research interest for RAN slicing and its impact on radio resource management (RRM). On that \cite{sallent2017radio} and \cite{vo2018slicing} provide interesting solutions for efficient resource management and orchestration.
From the perspective of slicing admission strategy optimization, the methods reported in~\cite{bega_TMC2019,sciancalepore2017slice,han2018slice} can be worthwhile to refer. A dynamic resource controller for vRANs based on deep reinforcement learning is presented in~\cite{vrain2019}, where the authors also showed a real implementation over different platforms. The authors of~\cite{GuoTVT_2019} introduces a novel framework for RAN slicing by showing performance requirements in terms of the required number of resources per deadline interval.
Although all these works only consider a binary decision mechanism where declined requests simply vanish instead of being served after a delay, the algorithms deployed by them to solve ILP problems will inspire future development of model-less heuristic strategy optimizers for the proposed multi-queue slice admission controller.

SlaaS shall be considered as a specific type of public cloud environment, where service sessions can be categorized into multiple types with significantly heterogeneous resource demands. Queuing theory has been widely applied for cloud computing services to model the statistics of service demand and delivered quality of service (QoS), such as~\cite{vilaplana2014queuing} and \cite{chang2016modeling}. Especially, service schedulers with heterogeneous queues for different service types are discussed in~\cite{li2017qos} and~\cite{guo2018optimal}. In addition, \cite{bonald_infocom17} relates to multi-resource sharing between flows with heterogeneous requirements providing a convergence proof. These models provide valuable reference views in addition to the model proposed in this paper. 
Finally, balking and reneging behavior of impatient clients in queuing systems are extensively studied in~\cite{bocquet2005queueing,yue2008waiting}.

Differing from the aforementioned works wherein a ``strategy'' usually represents the decision as a function of the system state, our study proposes a novel mechanism of multi-queuing slice admission control where the slicing strategy represents the MNO's preference of slice types in different system states. Besides, out paper also considers impatient tenants, which, from the best of our knowledge, has never been investigated in SlaaS environments.

\section{Conclusion}
\label{sect:concl}
The network slicing paradigm is expected to play a key-role in next generation networks design. However, devising an admission control solution that takes into account complex network tenants behaviors involves a large number of challenges. 

In this paper, we have proposed a multiservice-based network slicing controller that automatically accounts for tenants waiting to get their network slices request granted given certain request frequency and patience characteristics. Our results show that $i$) unexpected tenants behaviors may be modeled with advanced admission control policies, $ii$) the decisions of rational impatient tenants can be mapped onto classical queuing-theoretic models comprising balk and renege parameters and $iii$) numerical simulations closely follow the exponential balking/reneging models derived.

\appendix[The steady state probability of queue with exponential balking and reneging]\label{appendix}
Consider the queue of type-$n$ requests where the request arriving rate is $\lambda_n$, the request serving rate is $\mu_n$, the reneging factor is $\alpha_n$ and the balking factor is $1-e^{-\beta_nl/\mu_n}$. Let $p_n(l,t)$ denote the transient probability the queue contains $l$ requests at the time instant $t$, we can write the transition equations of the dynamic queue state:
\begin{align}
	\frac{\partial p_n'(0,t)}{\partial t}=&-\lambda_n p_n(0,t)+\mu_n p_n(1,t),\\
	\frac{\partial p_n'(0,t)}{\partial t}=&-\left(\lambda_ne^{-\beta_n/\mu_n}+\mu_n\right)p_n(1,t)+\lambda_n p_n(0,t)\\\nonumber
	&+(\mu_n+\alpha_n) p_n(2,t),\\
	\frac{\partial p_n'(0,t)}{\partial t}=&-\left(\lambda_ne^{-\beta_nl/\mu_n}+\mu_n+(l-1)\alpha_n\right)p_n(l,t)\nonumber\\
	&+\lambda_n e^{-\beta_n(l-1)/\mu_n} p_n(l-1,t)\\
	&+(\mu_n+l\alpha_n) p_n(l+1,t),\quad l\in\{3,4,5,\dots\}.\nonumber
\end{align}
Let $\gamma_n=\mu_n/\alpha_n$, $\kappa_{n}(l)=\lambda_ne^{-\gamma_nl/\mu_n}$, the steady-state equations are therefore
\begin{align}
	0=&-\kappa_{n}(0)p_n(0)+\gamma_np_n(1)\\
	0=&-(\kappa_{n}(1)+\gamma_n)p_n(1)+\kappa_{n}(0)p_n(0)+(\gamma_n+1)p_n(2)\\
	0=&-\left(\kappa_{n}(l)+\gamma_n+l-1\right)p_n(l)+\kappa_{n}(l-1)p_n(l-1)\\\nonumber
	&+(\gamma_n+l)p_n(l+1), l\in\{3,4,5,\dots\}
\end{align}
From the steady-state equations we have
\begin{equation}
	\begin{split}
		p_n(l)&=\frac{\kappa_n(l-1)}{\gamma_n+l-1}p_n(l-1)=p_n(0)\prod\nolimits_{i=1}^l\frac{\kappa_n(i)}{\gamma_n+i}\\
		=&p_n(0)\frac{\lambda_ne ^{-i\beta_n/\mu_n}}{\mu_n+i\alpha_n},\qquad l\in\mathbb{N}^+.
	\end{split}
\end{equation}
Knowing that $\sum\nolimits_{l=0}^{+\infty}p_n(l)=1$,  $p_n(0)$ can be calculated as
\begin{equation}
	p_n(0)=1\left/\left(1+\sum\nolimits_{l=1}^{+\infty}\prod\nolimits_{i=1}^l\frac{\lambda_ne^{-i\beta_n/\mu_n}}{\mu_n+i\alpha_n}\right)\right..
\end{equation}

\section*{Acknowledgment}
This work has been partially funded by the EU H2020 project 5G-CARMEN under Grant Agreement 825012.

%

%
%



\bibliographystyle{IEEEtran}
\bibliography{references}

 %
 	\begin{IEEEbiography}[{\includegraphics[width=1in,height=1.25in,clip,keepaspectratio]{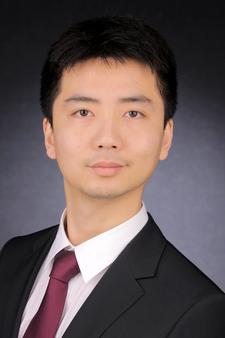}}]{Bin Han} (M'15) received in 2009 his B.E. degree from Shanghai Jiao Tong University, M.Sc. in 2012 from Technische Universit\"at Darmstadt, and in 2016 the Ph.D. (Dr.-Ing.) degree from Kalsruher Institut f\"ur Technologie. Since July 2016 he has been with Technische Universit\"at Kaiserslautern, researching in the broad area of wireless networks and signal processing, with recent special focus on network slicing and timely information delivery.
 	\end{IEEEbiography}

 	\begin{IEEEbiography}[{\includegraphics[width=1in,height=1.25in,clip,keepaspectratio]{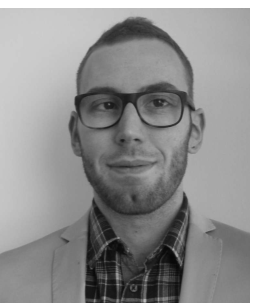}}]{Vincenzo Sciancalepore} (S'11--M'15--SM'19) received his M.Sc. degree in Telecommunications Engineering and Telematics Engineering in 2011 and 2012, respectively, whereas in 2015, he received a double Ph.D. degree. Currently, he is a senior 5G researcher at NEC Laboratories Europe GmbH in Heidelberg, focusing his activity on network virtualization and network slicing challenges. He is currently involved in the IEEE Emerging Technologies Committee leading the initiatives on SDN and NFV. He was also the recipient of the national award for the best Ph.D. thesis in the area of communication technologies (Wireless and Networking) issued by GTTI in 2015.
 	\end{IEEEbiography}

 	\begin{IEEEbiography}[{\includegraphics[width=1in,height=1.25in,clip,keepaspectratio]{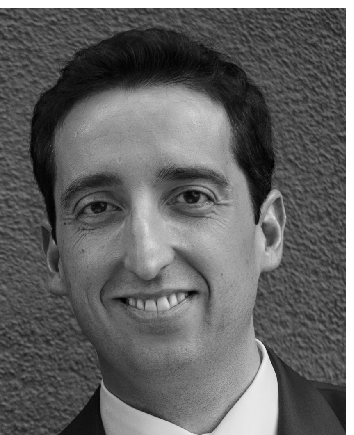}}]{Xavier Costa-P\'erez} (M'06--SM'18) is Head of 5G Networks R\&D and Deputy General Manager of the Security \& Networking Research Division at NEC Laboratories Europe. His team contributes to products roadmap evolution as well as to European Commission R\&D collaborative projects and received several awards for successful technology transfers. He received both his M.Sc. and Ph.D. degrees in Telecommunications from the Polytechnic University of Catalonia (UPC) in Barcelona and was the recipient of a national award for his Ph.D. thesis.
 	\end{IEEEbiography}

 	\begin{IEEEbiography}[{\includegraphics[width=1in,height=1.25in,clip,keepaspectratio]{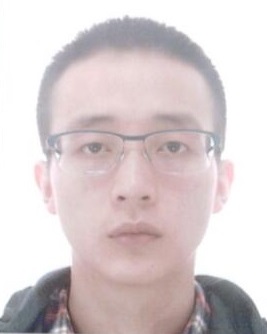}}]{Di Feng} received the M.A. in Economics from Yokohama National University and M.Res. in Economics from Universitat Aut\`onoma de Barcelona in 2016 and 2018, respectively. Currently, he is working for his Ph.D. degree in Economics at HEC - University of Lausanne. His research field covers decision theory, game theory and operations research, with a particular interest in market/mechanism design.
 	\end{IEEEbiography}

 	\begin{IEEEbiography}[{\includegraphics[width=1in,height=1.25in,clip,keepaspectratio]{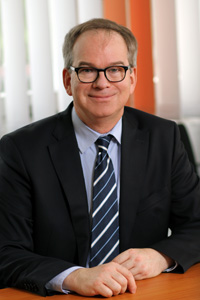}}]{Hans D. Schotten} (S'93--M'97) received the M.Sc. (Dipl.-Ing.) and Ph.D. (Dr.-Ing.) degrees in Electrical Engineering from the Aachen University of Technology RWTH, Germany in 1990 and 1997, respectively. Since August 2007, he has been full professor and head of the Institute of Wireless Communication at the University of Kaiserslautern. Since 2012, he has additionally been Scientific Director at the German Research Center for Artificial Intelligence heading the ``Intelligent Networks'' department.
 	\end{IEEEbiography}


\end{document}